\documentclass[3p,number,sort&compress]{elsarticle}

\journal{Annals of Physics}

\usepackage{amsmath}
\usepackage{amsfonts}
\usepackage{bbold,yfonts}
\usepackage{graphicx}
\usepackage{amssymb,mathrsfs}
\usepackage{upgreek}

\newcommand{\bs}[1]{\boldsymbol{#1}}

\begin{document}

\begin{frontmatter}

\title{Hydrodynamic approach to many-body systems: exact conservation laws}


\author[kit,mifi]{Boris N. Narozhny}
\ead{boris.narozhny@kit.edu}

\address[kit]{Institut f\"ur Theorie der Kondensierten Materie,
  Karlsruher Institut f\"ur Technologie, 76131 Karlsruhe, Germany}
\address[mifi]{National Research Nuclear University MEPhI 
  (Moscow Engineering Physics Institute), 115409 Moscow, Russia}

\begin{abstract}
In this paper I present a pedagogical derivation of continuity
equations manifesting exact conservation laws in an interacting
electronic system based on the nonequilibrium Keldysh technique. The
purpose of this exercise is to lay the groundwork for extending the
hydrodynamic approach to electronic transport to strongly correlated
systems where the quasiparticle approximation and Boltzmann kinetic
theory fail. 
\end{abstract}


\end{frontmatter}


\section*{}

Electronic hydrodynamics has evolved into a fast paced field with
multiple experimental and theoretical groups working to uncover
observable signatures of hydrodynamic behavior of electronic systems
\cite{pg,me0,luc,rev} with the primary focus on transport properties.
In a ``realistic'' case of a weakly disordered conductor, hydrodynamic
equations encompass the conventional linear-response transport theory
describing both the uniform Ohmic current in macroscopic
(``infinite'') systems and the nonuniform viscous flows of charge and
energy in constricted (``mesoscopic'') geometries \cite{gurzhi,fl0,fal19}.

Similarly to the traditional transport theory, hydrodynamic equations
can be derived from the kinetic (Boltzmann) equation describing a
system of weakly interacting quasiparticles \cite{me1}. At the semiclassical
level, one often relies on the ``scattering time approximation''
(typically used to describe Drude-like transport phenomena
\cite{chai}) to simplify the collision integral. This approach can be
further extended to include quantum interference phenomena
\cite{rammer,zna} yielding the so-called ``quantum corrections'' to
the leading semiclassical behavior heralding the onset of
low-temperature localization \cite{rama}. These additional features
``correct'' the conductivity of the system, while the macroscopic
description of the current flow remains Ohmic. The low-temperature
Ohmic resistance is still determined by disorder, although
electron-electron interaction does affect the quantum corrections
\cite{aar,zna}. In contrast, the hydrodynamic behavior is dominated by
electron-electron interaction determining the viscosity coefficient
\cite{me2}.

Systems dominated by electron-electron interactions, e.g., strongly
correlated systems, ``strange metals'', etc., remain a formidable
challenge for several decades. In simple terms, the difficulty lies in
the failure of the quasiparticle approach \cite{zaa15}. Moreover, even
if quasiparticles could be defined the semiclassical kinetic approach
may fail in multi-component systems with non-Abelian degrees of
freedom (e.g., spin or isospin) due to the purely quantum nature of
the latter \cite{aka}. It is then highly desirable to develop a
macroscopic theory of electronic transport in strongly interacting
systems without reliance on the quasiparticle paradigm and
semiclassical approximation which is the ultimate motivation for this
work.

In classical systems such a macroscopic theory is hydrodynamics.
Indeed, the Navier-Stokes equation \cite{navier,stokes,stokes51,dau6}
equally well describes water and air flows, while the Boltzmann
kinetic theory allowing one to derive this equation \cite{dau10} is
only justified for a dilute gas. The apparent universality of the
hydrodynamic theory can be attributed to two points: (i) the long
time, long distance behavior is often assumed to be independent of the
details of short-distance scattering processes, and (ii) the
conservation laws that are the basis of hydrodynamics are equally
applicable to all systems with the same symmetries.

Generalizing the hydrodynamic approach to systems beyond conventional
fluids, one may consider it in a broader sense meaning of a
long-wavelength theory of small perturbations relative to an
equilibrium state \cite{chai}. This way both the conventional
hydrodynamics and diffusion \cite{chai,df2} could be discussed on
equal footing. The difference between the two behaviors is momentum
conservation which is assumed in hydrodynamics and is broken in
diffusive systems. In solids, electronic momentum is never truly
conserved (it can be lost due to scattering off impurities, phonons,
etc.). However, in ultra-pure materials it may be possible to find an
intermediate temperature range where electron-electron interaction is
the dominant scattering process \cite{me0,luc,rev} as reflected by the
hierarchy of typical time scales $\tau_{ee}\ll\tau_{\rm
  dis},\tau_{e-ph},...$ (using self-evident notations). Then it could
be reasonable to neglect processes that do not conserve momentum, at
least as the ``0-th'' approximation describing the ``ideal fluid'' by
means of macroscopic (differential) equations essentially generalizing
\cite{me0,me1} the Euler's equation \cite{dau6}. The non-conserving
processes (electron-impurity or electron-phonon coupling) can then be
included perturbatively \cite{me0,luc,rev}.

The general problem of fermions with momentum-conserving interaction
has been one of the most popular in many body physics. Most generally,
the system is described by a Hamiltonian comprising the one-particle
(``free'') and ``interaction'' parts
\begin{subequations}
\label{H}
\begin{equation}
\widehat{H} = \widehat{H}_0 + \widehat{H}_{\rm int}.
\end{equation}
The one-particle contribution can typically be separated into two contributions
\begin{equation}
\label{h0}
\widehat{H}_0 =
\!\int\! d^d r_1 \hat\psi^\dagger(\bs{r}_1) \widehat{\cal H}^{(0)}_1 \hat\psi(\bs{r}_1),
\qquad
\widehat{\cal H}^{(0)}_1=\widehat{\cal K}_1 + U_1, 
\end{equation}
with $\widehat{\cal K}_1$ representing the ``kinetic energy''
(possibly including multiple bands, spin-orbit interaction, etc.;
without loss of generality all additional quantum numbers are
suppressed throughout this paper) and $U_1$ being the one-particle
potential [the subscript ``1'' refers to the set of quantum
numbers of the field $\hat\psi^\dagger(\bs{r}_1)$]. The interaction
term is assumed to be translationally invariant (hence, momentum-conserving)
\begin{equation}
\label{hint}
\widehat{H}_{\rm int} = \frac{1}{2}
\!\int\! d^d r_1 d^d r_2 \hat\psi^\dagger(\bs{r}_1)\hat\psi^\dagger(\bs{r}_2)
V(\bs{r}_1\!-\!\bs{r}_2) \hat\psi(\bs{r}_2)\hat\psi(\bs{r}_1).
\end{equation}
\end{subequations}
The general problem represented by the Hamiltonian (\ref{H}) cannot be
solved exactly. However, the conservation laws of particle number
(charge), energy, and momentum are exact.

In this paper, I explore the emergence of exact conservation laws in
the by now standard field-theoretic approach to nonequilibrium
systems, the Keldysh technique \cite{rammer,kam}. This issue has been
already extensively discussed in literature on general many-body
theory \cite{schwing,bk,baym} and nuclear physics
\cite{ivanov99,ivanov,knoll} establishing the integral relations
expressing the global symmetries of the system. The present paper
explores a somewhat different angle. I am interested in ``deriving''
the {\it local} continuity equations manifesting the conservation laws
that are the starting point of the hydrodynamic approach (e.g.,
conservation of the particle number, energy, and momentum). The point
is to express the macroscopic currents and densities in the most
general form (i.e., in terms of the exact quantities involved in the
diagrammatic technique including Green's functions, self-energies,
etc) allowing for a straightforward generalization to specific
condensed matter system including multiple bands and spin-orbit
interaction. The requirement of the ``exact'' validity of the
continuity equations leads to general relations involving the
self-energies and Green's functions. These relations are satisfied by
the exact functions and serve as constraints on their approximate
forms. At the same time, these relations provide a blueprint for
including additional, non-conserving terms to the Hamiltonian (e.g.,
electron-impurity or electron-phonon scattering) leading to weak decay
contributions to the resulting macroscopic equations
\cite{me0,hydro0,meig1}. Finally, I compare the obtained expressions
with those appearing as a result of the approximations leading to the
kinetic equation (semiclassical or quantum) as an intermediate
step. The ultimate goal of this work is to establish a hydrodynamic
framework that does not rely on the quasiparticle paradigm (avoiding
the kinetic equation and specifically the concept of the semiclassical
distribution function) and hence could be useful for describing
systems where quasiparticles are overdamped or altogether absent.

\section{Equations of motion}
\label{ce0}

In this paper I consider the conservation laws using the
nonequilibrium Keldysh technique following Ref.~\cite{baym}. The
notations for the Keldysh Green's functions and general relations
between them are summarized in~\ref{appa}, for a more detailed account
of the Keldysh technique see Refs.~\cite{rammer,kam}.

All Green's functions are defined in terms of the Heisenberg field
operators and therefore it is important to review the equations of
motion governing their dynamics. Starting with the standard
quantum-mechanical definition of the time derivative,
\begin{subequations}
\label{eqmo}
\begin{equation}
\label{eqmo0}
i\frac{\partial}{\partial t} \hat\psi = \left[\hat\psi, \widehat{H}\right].
\end{equation}
one finds \cite{schwing}
\begin{equation}
\label{eqmo1}
i\frac{\partial}{\partial t} \hat\psi(\bs{r},t) - \widehat{\cal H}^{(0)}  \hat\psi(\bs{r},t)
=
\int d^dr' \hat\psi^\dagger(\bs{r}',t) V(\bs{r}\!-\!\bs{r}') \hat\psi(\bs{r}',t) \hat\psi(\bs{r},t).
\end{equation}
\end{subequations}
Multiplying Eq.~(\ref{eqmo1}) by $i\hat\psi^\dagger$ from the left and
taking the thermodynamic average, one arrives at the Dyson's equation
for the ``12'' component of the Keldysh Green's function, see
Eq.~(\ref{dyeq0}), but with the right-hand side (RHS) expressed
explicitly in terms of the interaction potential
\begin{subequations}
\label{dyeq120}
\begin{equation}
\label{dyeq12v}
i\frac{\partial}{\partial t_1}G^{12}_{1,2} \!-\! \widehat{\cal H}^{(0)}_1 G^{12}_{1,2}=
i\!\int\! d^dr_3
\left\langle\hat\psi^\dagger(\bs{r}_2, t_2) \hat\psi^\dagger(\bs{r}_3,t_1) 
V(\bs{r}_1\!-\!\bs{r}_3) 
\hat\psi(\bs{r}_3,t_1) \hat\psi(\bs{r}_1, t_1)\right\rangle.
\end{equation}
Here and throughout the paper I use the short-hand notation: 
$G^{12}_{1,2}=G^{12}(\bs{r}_1,t_1; \bs{r}_2,t_2)$.

In contrast, the Dyson's equation is expressed in terms of the
self-energy, see Eq.~(\ref{dyeq0})
\begin{equation}
i\frac{\partial}{\partial t_1} G^{12}_{1,2}\!-\! \widehat{\cal H}^{(0)}_1 G^{12}_{1,2}=
\!\int\! d3 \left[ \Sigma^{11}_{1,3} G^{12}_{3,2} - \Sigma^{12}_{1,3} G^{22}_{3,2} \right]\!,
\end{equation}
where $d3=d^dr_3dt_3$. Alternatively [using Eqs.~(\ref{grak}) and the
  similar relation for the self-energy]
\begin{equation}
\label{dyeq12}
i\frac{\partial}{\partial t_1} G^{12}_{1,2}\!-\! \widehat{\cal H}^{(0)}_1 G^{12}_{1,2}=
\!\int\! d3 \, \Xi(1,2;3),
\qquad
\Xi(1,2;3) = \Sigma^R_{1,3} G^{12}_{3,2} \!+\! \Sigma^{12}_{1,3} G^A_{3,2} .
\end{equation}
Comparing Eqs.~(\ref{dyeq12v}) and (\ref{dyeq12}), one arrives at the identity
\begin{equation}
\label{hintid}
i\!\int\! d^dr_3
\left\langle\hat\psi^\dagger(\bs{r}_2, t_2) \hat\psi^\dagger(\bs{r}_3,t_1) 
V(\bs{r}_1\!-\!\bs{r}_3) 
\hat\psi(\bs{r}_3,t_1) \hat\psi(\bs{r}_1, t_1)\right\rangle
=
\!\int\! d3 \, \Xi(1,2;3),
\end{equation}
\end{subequations}
relating the two-particle Green's function in the RHS of
Eq.~(\ref{dyeq12v}) to the single-particle quantities in the RHS of
Eq.~(\ref{dyeq12}).

In what follows, I will also use the equation of motion for
$\hat\psi^\dagger$
\begin{equation}
\label{eqmo2}
i\frac{\partial}{\partial t} \hat\psi^\dagger(\bs{r},t) 
+ \hat\psi^\dagger(\bs{r},t)\widehat{\cal H}^{(0),\dagger}  
=
-\int d^dr' 
\hat\psi^\dagger(\bs{r},t) V(\bs{r}\!-\!\bs{r}') \hat\psi^\dagger(\bs{r}',t) \hat\psi(\bs{r}',t).
\end{equation}
Here $\widehat{\cal H}^{(0),\dagger}$ is the conjugate operator with
any gradients acting on the coordinate dependence to the
left. Multiplying this equation by $i\hat\psi$ from the right one
finds
\begin{subequations}
\label{dyeq120cc}
\begin{equation}
\label{dyeq12vcc}
i\frac{\partial}{\partial t_2}G^{12}_{1,2} \!+\! \widehat{\cal H}^{(0),\dagger}_2 G^{12}_{1,2}=
-i\!\int\! d^dr_3
\left\langle\hat\psi^\dagger(\bs{r}_2, t_2) \hat\psi^\dagger(\bs{r}_3,t_2) 
V(\bs{r}_1\!-\!\bs{r}_3) 
\hat\psi(\bs{r}_3,t_2) \hat\psi(\bs{r}_1, t_1)\right\rangle.
\end{equation}
On the other hand, conjugating the Dyson's equation (\ref{dyeq12})
[see Eqs.~(\ref{gfcon}), (\ref{selfcon}), (\ref{red1}),
  (\ref{selfen1}), and (\ref{gfcon12})] and changing the variables
$1\leftrightarrow2$ one arrives at
\begin{equation}
\label{dyeq12cc}
i\frac{\partial}{\partial t_2} G^{12}_{1,2} \!+\! \widehat{\cal H}^{(0),\dagger}_2 G^{12}_{1,2}=
\!\int\! d3 \, \Xi^*(2,1;3) =
- \!\int\! d3 \left[ G^R_{1,3} \Sigma^{12}_{3,2} + G^{12}_{1,3} \Sigma^A_{3,2} \right].
\end{equation}
\end{subequations}
Comparing Eqs.~(\ref{dyeq12vcc}) and (\ref{dyeq12cc}) yields the
conjugate form of the identity (\ref{hintid}).

\section{Continuity equation}
\label{ce1}

Consider now the usual continuity equation
\begin{equation}
\label{coneq}
\frac{\partial n}{\partial t} + \bs{\nabla}\!\cdot\!\bs{j} = 0.
\end{equation}
The continuity equation itself is well-known and does not need another
derivation. This equation represents gauge invariance (i.e., the
particle number conservation or charge conservation), the symmetry
that is typically assumed to be exact for all condensed matter systems
(apart from the special case of superconductivity where this issue is
more subtle, see Ref.~\cite{rammer}) and hence is independent of the
particular form of the Hamiltonian. The purpose of this section is to
introduce notations for the particle number density, $n$, and the
current, $\bs{j}$, and establish the constraint imposed on the
self-energy by gauge invariance.

\subsection{Continuity equation at the operator level}

The particle number can be defined in the standard way using
electronic field operators 
\begin{subequations}
\label{ndef}
\begin{equation}
\label{n0}
\hat n(\bs{r}, t) = \hat\psi^\dagger(\bs{r}, t) \hat\psi(\bs{r}, t),
\qquad
n(\bs{r}, t) = \left\langle \hat n(\bs{r}, t) \right\rangle,
\end{equation}
or the Green's function [see Eq.~(\ref{g12})]
\begin{equation}
\label{ngf}
n_1 = -i G^{12}_{1,1}.
\end{equation}
\end{subequations}
The two definitions allow for two different derivations of the
continuity equation starting either with the equations of motion or
the Dyson's equations.

At the operator level, the continuity equation is just the equation of
motion for the density operator that can be obtained by combining the
two equations of motion (\ref{eqmo1}) and (\ref{eqmo2})
\begin{equation}
\label{eqmon}
\frac{\partial}{\partial t} \hat n(\bs{r}, t)
=
-i\hat\psi^\dagger(\bs{r},t)\widehat{\cal K} \hat\psi(\bs{r},t)
+i
\left[\widehat{\cal K}\hat\psi(\bs{r},t)\right]^\dagger \hat\psi(\bs{r},t)
=
-\bs{\nabla}\!\cdot\!\hat{\bs{j}}(\bs{r}, t),
\end{equation}
where the last step defines the current operator. The interaction
potential does not appear in Eq.~(\ref{eqmon}) due to the standard
commutation relations.

\subsection{Continuity equation and the Keldysh Green's functions}

Combining Eq.~(\ref{dyeq12}) with Eq.~(\ref{dyeq12cc}) yields
a Kadanoff-Baym equation \cite{bk}
\begin{eqnarray}
\label{dyeq12dif}
&&
\left[i\frac{\partial}{\partial t_1} \!-\! \widehat{\cal H}^{(0)}_1
+i\frac{\partial}{\partial t_2} \!+\! \widehat{\cal H}^{(0),\dagger}_2\right]\! G^{12}_{1,2}=
\!\int\! d3 \left[\, \Xi(1,2;3) + \Xi^*(2,1;3) \right] \\
&&
\nonumber\\
&&
\qquad\qquad
=
\!\int\! d3 \left[ 
\Sigma^R_{1,3} G^{12}_{3,2} + \Sigma^{12}_{1,3} G^A_{3,2}-
G^R_{1,3} \Sigma^{12}_{3,2} - G^{12}_{1,3} \Sigma^A_{3,2} 
\right]\!.
\nonumber
\end{eqnarray}
Comparing the time derivative terms in Eq.~(\ref{dyeq12dif}) to the
definition of the particle density, see Eq.~(\ref{ndef}), one notices
the relation 
\[
\left.\left[i\frac{\partial}{\partial t_1} \!+\! 
i\frac{\partial}{\partial t_2}\right]\! G^{12}_{1,2}
\right|_{2\rightarrow1} =
-\frac{\partial n_1}{\partial t_1} .
\]
Now it becomes clear that in the limit $2\rightarrow1$
Eq.~(\ref{dyeq12dif}) can be written in the form of the conventional
continuity equation (\ref{coneq}) where the divergence of the current
is determined by the single-particle Hamiltonian
\begin{equation}
\label{jd0}
\bs{\nabla}_1\!\cdot\!\bs{j}_1 = 
\left.\left[\widehat{\cal K}_1-\widehat{\cal K}^\dagger_2\right]\! G^{12}_{1,2}
\right|_{2\rightarrow1},
\end{equation}
while the self-energy satisfies the condition
\begin{equation}
\label{id10}
\int\! d3 \left[ 
\Sigma^R_{1,3} G^{12}_{3,1} + \Sigma^{12}_{1,3} G^A_{3,1}-
G^R_{1,3} \Sigma^{12}_{3,1} - G^{12}_{1,3} \Sigma^A_{3,1} 
\right] =0.
\end{equation}
The latter identity is satisfied by the exact self-energy and Green's
function and hence represents a constraint on any approximate
expressions. In fact, the identity (\ref{id10}) can be derived
independently, following Refs.~\cite{bk,baym}, where the idea of
``conserving approximations'' was first suggested.

\subsection{Conserving approximations}

The need for a ``conserving approximation'' arises from the apparent
arbitrariness of the diagrammatic perturbation theory. Indeed, it may
not be clear ``a priori'' that a given approximation for the
self-energy satisfies the exact conservation laws of the system (given
that this is certainly not the case for at least some individual
diagrams; of course, any practitioner of the diagrammatic perturbation
theory would make sure that the calculation does not violate gauge
invariance, although this might involve certain technical difficulties
-- the point of a ``conserving approximation'' is that the
conservation laws are satisfied automatically without any need for
special care). Baym suggested the self-consistent procedure where one
starts with the Luttinger-Ward functional
\begin{equation}
\label{phidef}
\Phi[\check G] = \left[\ln\left\langle\widehat{\cal S}_C\right\rangle\right]_{\rm sk}
=
\left[\left\langle\widehat{\cal S}_C\right\rangle-1\right]_{\rm sk},
\quad
\Phi^*=\Phi,
\end{equation}
where the subscript ``{\rm sk}'' indicates that only skeleton diagrams
(i.e. diagrams without self-energy insertions and with all Green's
functions replaced by full Green's functions) are to be
retained. Moreover, the logarithm amounts to retaining only the
connected diagrams. The important property of the functional is that
the exact self-energy can be obtained by the variation
\begin{equation}
\label{sephi}
\Sigma^{ij}_{1,2} = - (-1)^{i+j} \frac{\delta\Phi}{\delta G^{ji}_{2,1}}.
\end{equation}
The ``self-consistent'' perturbation theory comprises an expansion of
the functional $\Phi$ and a solution for $\check G$ and $\check\Sigma$
using the Dyson's equation (\ref{dyeqk}) and Eq.~(\ref{sephi}). The
latter step is self-consistent in the sense that Eq.~(\ref{dyeqk})
determines the Green's function in terms of the self-energy and
Eq.~(\ref{sephi}) the other way around. The key point of the
self-consistent approach is that the resulting theory satisfies exact
conservation laws without any further approximation no matter how many
diagrams are retained in the expansion of the functional $\Phi$
\cite{baym}. The resulting approximations are known as
``$\Phi$-derivable''. While there can be many such approximations
(depending on the order to which $\Phi$ is expanded), all of them
respect the conservation laws.

\subsection{$\Phi$-derivable approximations and gauge invariance}

Applying a symmetry transformation to the exact Green's function leads
to a variation of the functional. To the leading order, the variation
$\delta\Phi$ is given by
\begin{equation}
\label{dphi}
\delta\Phi 
=
- \!\int\! d1d2 \,{\rm Tr}\, \check\tau_3 \check\Sigma_{1,2} \check\tau_3 \delta \check G_{2,1},
\end{equation}
which vanishes if the transformation corresponds to a true symmetry of
the Hamiltonian.

Consider a gauge transformation (cf. the same argument of
Ref.~\cite{baym} but using the Matsubara Green's functions) which
without loss of generality can be confined to the upper branch of the
Keldysh contour
\begin{subequations}
\label{gt1}
\begin{equation}
\check G_{2,1} \rightarrow e^{i\check\chi_2} \check G_{2,1} e^{-i\check\chi_1},
\quad
\check\chi = 
\begin{pmatrix}
\chi & 0 \cr
0 & 0
\end{pmatrix}.
\end{equation}
Expanding to the leading order in $\chi$ and taking into account the
matrix structure, one finds
\begin{equation}
\delta \check G_{2,1} = i \chi_2 \frac{1\!+\!\check\tau_3}{2} \check G_{2,1}
-
i \chi_1 \check G_{2,1} \frac{1\!+\!\check\tau_3}{2}
\end{equation}
\end{subequations}
Substituting this expression into Eq.~(\ref{dphi}) and requiring that
the functional is invariant under the gauge transformation (since it
is composed of closed particle lines) one finds (using the cyclic
property of the trace in each term separately)
\begin{subequations}
\label{grtr}
\begin{eqnarray}
\int\!d1\chi_1\int\!d2 \, {\rm Tr} \frac{1\!+\!\check\tau_3}{2}
\left[ 
\check\tau_3\check\Sigma_{1,2}\check\tau_3\check G_{2,1}
-
\check G_{1,2}\check\tau_3\check\Sigma_{2,1}\check\tau_3
\right]=0.
\end{eqnarray}
Evaluating the trace and taking into account arbitrariness of
$\chi_1$ one arrives at the identity
\begin{eqnarray}
\label{id1}
\int\!d2
\left[ 
\Sigma^{11}_{1,2}G^{11}_{2,1} - \Sigma^{12}_{1,2}G^{21}_{2,1}
-G^{11}_{1,2}\Sigma^{11}_{2,1} + G^{12}_{1,2}\Sigma^{21}_{2,1}
\right]=0,
\end{eqnarray}
which is a manifestation of gauge invariance.

Now, substituting Eqs.~(\ref{grak}) into Eq.~(\ref{id1}), one finds
\begin{eqnarray}
\label{sigg}
\Sigma^{11}_{1,2}G^{11}_{2,1} - \Sigma^{12}_{1,2}G^{21}_{2,1}
=
\Sigma^R_{1,2} G^{12}_{2,1} + \Sigma^{12}_{1,2} G^A_{2,1}
+ \Sigma^R_{1,2} G^R_{2,1}.
\end{eqnarray}
Comparing Eqs.~(\ref{id10}) and (\ref{id1}) I now conclude that in the limit
$2\rightarrow1$ the integral in the right-hand side of
Eq.~(\ref{dyeq12dif}) takes the form
\begin{equation}
\label{grgr}
\int\! d3 \left[  \Sigma^R_{1,3} G^R_{3,1} - G^R_{1,3}\Sigma^R_{3,1} \right] = 0.
\end{equation}
\end{subequations}
This expression vanishes for the following reasons: (i) the
self-energy has the same causality structure as the Green's function
\cite{kam}, therefore for any $t_1\ne t_3$ the product
$\Sigma^R_{1,3}G^R_{3,1}$ vanishes; (ii) in the limit
$t_3\rightarrow t_1$ the retarded Green's function has the form
\begin{equation}
\label{gr11}
G^R(\bs{r}_1, \bs{r}_3; t_1=t_3+0)=-i\delta(\bs{r}_1-\bs{r}_3),
\end{equation}
so that even if the self-energy had a non-zero diagonal value
$\Sigma^R(1,1)$ it would be the same in both terms and hence canceled
in the difference. As a result, the identity (\ref{id10}) follows from
Eq.~(\ref{id1}). The above argument represents a proof of
Eq.~(\ref{id10}) and, by extension, confirms that the continuity
equation (\ref{coneq}) is consistent with the Keldysh approach
(exactly or within a $\Phi$-derivable approximation).

\subsection{Summary}

To summarize this section, the continuity equation (\ref{coneq})
manifesting particle number conservation follows from the Heisenberg
equations of motion due to the ``density-density'' interaction, see
Eq.~(\ref{hint}). It is fully preserved in the microscopic Keldysh
approach (either while using the exact Green's functions or within a
$\Phi$-derivable -- or any other conserving -- approximation). At the
same time, the continuity equation is satisfied within the kinetic
theory (that can be derived from the same microscopic theory using a
series of approximations, see Ref.~\cite{rammer} and
Section~\ref{sec:kinur}) as well. Given that particle number
conservation is the exact symmetry, the continuity equation is valid
independently of any (correctly applied) approximation. Specifically,
the arguments presented here do not rely on either the quasiparticle
and semiclassical approximations typically assumed to derive the
kinetic equation or otherwise describe conventional metals and
semiconductors.

\section{Momentum conservation}
\label{sec:momcon}

Consider now translational invariance. This is the crucial symmetry in
conventional hydrodynamics, where the Navier-Stokes equation
\cite{dau6} is a direct consequence of momentum conservation.

Since the Hamiltonian is explicitly translationally invariant, one
should be able to express momentum conservation by means of the
continuity equation for the momentum density, $\bs{g}$, similarly to
Eq.~(\ref{coneq})
\begin{equation}
\label{mconeq}
\frac{\partial g_\alpha}{\partial t}  +
\nabla^\beta\tau_{\beta\alpha} = 0,
\end{equation}
without any additional derivation (here $\tau_{\alpha\beta}$ is the
momentum flux -- or stress -- tensor). However, there are well
documented difficulties along the way \cite{schwing,bk,baym},
primarily for long-ranged interactions.

Within the kinetic theory, one may derive the Navier-Stokes equation
by multiplying the kinetic equation by momentum and integrating over
all single-particle states \cite{dau10,me0,me1}. The equivalent
procedure at the microscopic level is to apply the momentum operator
to the Dyson equations (\ref{dyeq12}) and (\ref{dyeq12cc}) followed by
the evaluating their sum in the limit ${2\rightarrow1}$. In the
resulting equation [similar to Eq.~(\ref{dyeq12dif})], the time
derivative terms combine into the time derivative of momentum density,
while the rest should comprise the spatial derivatives yielding the
divergence of the momentum flux tensor and the ``collision integral''
terms vanishing in the limit ${2\rightarrow1}$, essentially repeating
the above calculation leading to the continuity equation
(\ref{coneq}).

\subsection{Momentum density}

The ``momentum operator'' mentioned above is the differential operator
allowing one to define the momentum density. In quantum field theory,
however, the definition of such operator is not unique
\cite{dau2,itz}. The reason is that only the {\it total momentum} of
the system is well-defined. While it can be expressed as a volume
integral over the momentum density, that integral remains unchanged if
any contribution representing a surface term is added to the
integrand. This freedom can be used to bring the stress tensor to a
symmetric form typically assumed in calculations of the viscosity
tensor \cite{read,brad,poli16,julia1}. Taking into account the
possible additional terms (important for non rotationally invariant
systems \cite{julia1}), the most general form of the momentum density
can be written as [cf. Eqs.~(\ref{ndef})]
\begin{subequations}
\label{mndef}
\begin{equation}
\label{mn0}
\bs{g}(\bs{r}, t) = \frac{1}{2}
\left\langle 
\hat\psi^\dagger(\bs{r}, t) \hat{\bs{p}} \hat\psi(\bs{r}, t) 
+
\left[\hat{\bs{p}}^\dagger \hat\psi^\dagger(\bs{r}, t)\right] \hat\psi(\bs{r}, t) 
\right\rangle,
\end{equation}
where $\hat{\bs{p}}$ is the momentum operator appropriate for the
system in question. Alternatively, the momentum density can be
expressed in terms of the Green's function Eq.~(\ref{g12})
\begin{equation}
\label{mngf}
\bs{g}_1 = -\frac{i}{2} \left[\hat{\bs{p}}_1+\hat{\bs{p}}^\dagger_2\right] G^{12}_{1,2}\Big|_{2\rightarrow1}.
\end{equation}
\end{subequations}
In conventional systems with the parabolic spectrum the momentum
operator has the usual form ${\hat{\bs{p}}=-i\bs{\nabla}}$ and the
resulting momentum density (\ref{mngf}) is proportional to the
particle number current (\ref{jd0}). This proportionality does not
hold in general (e.g., in the case of Dirac fermions in graphene
\cite{me0,me1,rev}).

\subsection{Global momentum conservation}

Following the above procedure, I now apply the operator
$-(i/2)[\hat{\bs{p}}_1+\hat{\bs{p}}^\dagger_2]$ to the Dyson equations
(\ref{dyeq12}) and (\ref{dyeq12cc}), sum up the results, and take
the limit ${2\rightarrow1}$. This yields
\begin{subequations}
\label{momcondeq}
\begin{equation}
\label{dyeq12difm}
\frac{\partial g^i_1}{\partial t_1} 
+ \left.n_1\,\frac{i\hat{p}^i_1U_1\!-\!i\hat{p}^{\dagger, i}_2U_2}{2}  \right|_{2\rightarrow1}
+ \nabla^j_1 \tau_0^{ji}(1)
= C^i_1,
\end{equation}
where 
\begin{equation}
\label{pi0}
\nabla^j_1 \tau_0^{ji}(1) 
= \left.\frac{\hat{p}^i_1\!+\!\hat{p}^{\dagger, i}_2}{2}
\left[\widehat{\cal K}_1-\widehat{\cal K}^\dagger_2\right]\! G^{12}_{1,2}
\right|_{2\rightarrow1},
\end{equation}
and [see Eqs.~(\ref{dyeq120}) and (\ref{dyeq12cc})]
\begin{equation}
\label{cdef}
\bs{C}_1=
-\left.
\frac{\hat{\bs{p}}_1\!+\!\hat{\bs{p}}^{\dagger}_2}{2}\!\int\! d3 \Big[ \,
\Xi(1,2; 3) + \Xi^*(2,1; 3)
\Big]\right|_{2\rightarrow1}\!\!.
\end{equation}
\end{subequations}
Conservation of total momentum can be demonstrated by integrating
Eq.~(\ref{dyeq12difm}) over the system volume and requiring that the
volume integral of the RHS vanishes
\begin{equation}
\label{momcontot}
\frac{\partial}{\partial t_1}\int  d^dr_1 \, \bs{g}_1
+
\int\!d^dr_1 
\left. n_1 \,\frac{i\hat{\bs{p}}_1U_1\!-\!i\hat{\bs{p}}^{\dagger}_2U_2}{2}\right|_{2\rightarrow1} 
= 0,
\qquad
\int\!d^dr_1 \,\bs{C}_1=0.
\end{equation}

The integral nature of the conservation law is consistent with the fact
that it is the total momentum of the system that is well defined and
conserved. A local momentum flux might not be well defined if
interactions are long ranged.

\subsection{$\Phi$-derivable approximations and translational invariance}

The last equality in Eq.~(\ref{momcontot}) represents a constraint on
the approximate self energies and Green's functions and can be proven
similarly to Eq.~(\ref{id10}). Consider a coordinate shift with the
operator
\begin{equation}
\label{t0}
\widehat{T}_{\bs{R}} = e^{i\bs{R}\cdot\hat{\bs{p}}}.
\end{equation}
Confining the shift to the upper branch of the contour in analogy with
Eq.~(\ref{gt1}), the transformation of the Green's function can be
expressed as
\begin{subequations}
\label{gt2}
\begin{equation}
\check G_{2,1} \rightarrow 
\frac{1\!+\!\check\tau_3}{2} 
\widehat{T}_{\bs{R}} \check G_{2,1} \widehat{T}^\dagger_{\bs{R}} \frac{1\!+\!\check\tau_3}{2}.
\end{equation}
To the leading order in $\bs{R}$, the variation of the Green's
function is given by
\begin{equation}
\delta \check G_{2,1} \!=\! 
i\frac{1\!+\!\check\tau_3}{2} \bs{R}(t_2)\hat{\bs{p}}_2\check G_{2,1}
-
i\bs{R}(t_1)\hat{\bs{p}}^\dagger_1\check G_{2,1} \frac{1\!+\!\check\tau_3}{2}.
\end{equation}
\end{subequations}
Substituting this expression into Eq.~(\ref{dphi}) and requiring that
the functional is invariant under the shift of coordinates (since only
the system boundaries are shifted) one finds (using the cyclic
property of the trace in each term separately)
\begin{eqnarray*}
\delta\Phi=i\int d1 \bs{R}(t_1) \frac{\hat{\bs{p}}_1\!+\!\hat{\bs{p}}^\dagger_3}{2}
\int d2 
\left.
\Big[ 
\Sigma^{11}_{1,2}G^{11}_{2,3} - \Sigma^{12}_{1,2}G^{21}_{2,3}
-G^{11}_{1,2}\Sigma^{11}_{2,3} + G^{12}_{1,2}\Sigma^{21}_{2,3}
\Big]\right|_{3\rightarrow1}\!=0,
\end{eqnarray*}
such that due to arbitrariness of $\bs{R}$ one arrives at
\begin{equation}
\label{id2}
\frac{\hat{\bs{p}}_1\!+\!\hat{\bs{p}}^\dagger_3}{2} \int d2
\left.
\Big[ 
\Sigma^{11}_{1,2}G^{11}_{2,3} - \Sigma^{12}_{1,2}G^{21}_{2,3}
-G^{11}_{1,2}\Sigma^{11}_{2,3} + G^{12}_{1,2}\Sigma^{21}_{2,3}
\Big]\right|_{3\rightarrow1}=0,
\end{equation}
which is a manifestation of translational invariance.

The expression in the square brackets in Eq.~(\ref{id2}) coincides
with that in Eq.~(\ref{id1}), while the corresponding combination of
the self-energies and Green's functions in Eq.~(\ref{cdef}) is the
same as in Eq.~(\ref{id10}). Given that the momentum operator does not
affect time dependence and hence causality, one can use the same
argument as in Section~\ref{ce1} concluding that the terms containing
products of two retarded (or two advanced) functions vanish in the
limit $2\rightarrow1$. Thus, the identity (\ref{id2}) proves the
second identity in Eq.~(\ref{momcontot}) and consequently, the
integral relation manifesting the momentum conservation.

\subsection{Momentum conservation in the local approximation}
\label{sec:momloc}

A local (``differential'') version of the momentum conservation law
can not be established without some degree of approximation
\cite{schwing,bk,baym}. Within the kinetic approach, the local
continuity equation for the momentum density, see Eq.~(\ref{mconeq})
is obtained by a straightforward integration of the kinetic equation
multiplied by momentum. This is possible because the distribution
function, the central quantity the kinetic theory, is already
local. In contrast, ``integrating'' the Kadanoff-Baym equation
(\ref{dyeq12dif}) leads to Eq.~(\ref{dyeq12difm}). This is not a
continuity equation since the quantity $\bs{C}$ in the RHS {\it is
  not a divergence}, see also Refs.~\cite{schwing,baym}.

The Kadanoff-Baym equation (\ref{dyeq12dif}) can also be derived using
the alternative form of the Dyson's equations, i.e.,
Eqs.~(\ref{dyeq12v}) and (\ref{dyeq12vcc}). This leads to the
expression for the quantity $\bs{C}$ in terms of the two-particle
Green's function [one could also use the identity (\ref{hintid}) in
  Eq.~(\ref{cdef})]
\begin{equation}
\label{c2}
\bs{C}_1 = -\frac{i}{2}\!\int\! d^d r_3 
\left.
\left[\hat{\bs{p}}_1 V(\bs{r}_1\!-\!\bs{r}_3) \!-\! \hat{\bs{p}}_2^\dagger V(\bs{r}_2\!-\!\bs{r}_3)
\right]\right|_{2\rightarrow1}\!
\left\langle\hat\psi_H^\dagger(\bs{r}_1, t_1) \hat\psi_H^\dagger(\bs{r}_3,t_1) 
\hat\psi_H(\bs{r}_3,t_1) \hat\psi_H(\bs{r}_1, t_1)\right\rangle\!.
\end{equation}
This form immediately proves that the volume integral of $\bs{C}$
vanishes, see Eq.~(\ref{momcontot}): indeed, the integrand is
antisymmetric, which reflects the third Newton's law \cite{schwing}.

The quantity $\bs{C}$ is not a divergence since the interaction
potential is nonlocal. However, for short-range interactions (e.g.,
for sufficiently screened Coulomb potential in solids) it is possible
to construct an effective local interaction stress tensor by
integrating $\bs{C}$ over a large enough volume \cite{schwing}
\begin{subequations}
\label{c1int}
\begin{equation}
\label{cint1}
\int\limits_V d^d r_1 \bs{C}_1 = 
- \int\limits_V d^d r_1 \int d^dr_3 \, \bs{c}_+(\bs{r}_1,\bs{r}_3),
\qquad
\bs{c}_+(\bs{r}_1,\bs{r}_3) = - \bs{c}_+(\bs{r}_3,\bs{r}_1),
\end{equation}
where $\bs{c}_+(\bs{r}_1,\bs{r}_3)$ is the integrand in $\bs{C}$,
which can be expressed in terms of single-particle functions due to
the identity (\ref{hintid})
\begin{eqnarray}
\label{cm}
\bs{c}_+(\bs{r}_1,\bs{r}_3) = 
\left.
\frac{\hat{\bs{p}}_1\!+\!\hat{\bs{p}}_2^\dagger}{2}
\int dt_3 
\Big[ \, \Xi(1,2; 3) + \Xi^*(2,1; 3) \Big]\right|_{2\rightarrow1}\!.
\end{eqnarray}
\end{subequations}
Since $\bs{c}_+(\bs{r}_1,\bs{r}_3)$ is antisymmetric, the integral
over any identical region in $\bs{r}_1$ and $\bs{r}_3$ vanishes. Thus
the coordinate $\bs{r}_3$ in Eq.~(\ref{cint1}) is effectively outside
of the volume $V$, while $\bs{r}_1$ is inside $V$ and the relative
coordinate, ${\bs{r}_{13}=\bs{r}_1-\bs{r}_3}$, is restricted by the
interaction range. Changing the integration variables in
Eq.~(\ref{cint1}) to $\bs{r}_{13}$ and $\bs{r}_3$, the integral takes
the form [where $V_{\bs{r}_1}$ indicates that the integration volume
  is $V$ in terms of the original variable $\bs{r}_1$ and
  ${\bs{R}_{13}=(\bs{r}_1\!+\!\bs{r}_3)/2}$]
\[
\int\limits_V d^d r_1 \int d^dr_3 \,\bs{c}_+(\bs{r}_1,\bs{r}_3)
=
\int\limits_{V_{\bs{r}_1}} d^d r_{13} d^d r_3 \,
\bs{c}_+(\bs{R}_{13}\!+\!\bs{r}_{13}/2,\bs{R}_{13}\!-\!\bs{r}_{13}/2).
\]
Martin and Schwinger \cite{schwing} introduced the hypothesis of
``local uniformity'' where expectation values of field operators
within a physically small region depend only on the relative
coordinate. Then for a fixed $\bs{r}_{13}$ the integration over
$\bs{r}_3$ is restricted to a shell of thickness
$\bs{n}\!\cdot\!\bs{r}_{13}$, where $\bs{n}$ is a unit vector normal
to the surface of $V$. Now the approximation of Ref.~\cite{schwing}
can be asserted by setting
${\bs{R}_{13}\approx\bs{r}_3\approx{const}}$ in that shell. The
integration measure over $\bs{r}_3$ can be replaced by
$-\bs{n}\!\cdot\!\bs{r}_{13} dS_3$ with the integral covering half the
volume in $\bs{r}_{13}$
\[
-\frac{1}{2} \!\int\! d^d r_{13} \!\int\! dS_3 (\bs{n}\!\cdot\!\bs{r}_{13})
\bs{c}_+(\bs{r}_3\!+\!\bs{r}_{13}/2,\bs{r}_3\!-\!\bs{r}_{13}/2)
=
-\frac{1}{2} \!\int\! dS_3 n^i\!\int\! d^d r_{13} r^i_{13}
\bs{c}_+(\bs{r}_3\!+\!\bs{r}_{13}/2,\bs{r}_3\!-\!\bs{r}_{13}/2).
\]
Now one can invoke the Euler's theorem and approximate the quantity
$\bs{C}$ by a divergence
\begin{equation}
\label{piint}
C^i(\bs{r})\approx -\nabla^j \tau^{ji}_{int},
\qquad
\tau^{ji}_{int} = -\frac{1}{2} \!\int\! d^d r_{13} r^j_{13} 
c^i_+(\bs{r}\!+\!\bs{r}_{13}/2,\bs{r}\!-\!\bs{r}_{13}/2).
\end{equation}
The quantity $\tau^{ji}_{int}$ represents the interaction contribution
to the stress-tensor.

Formally, one can arrive at Eq.~(\ref{piint}) by changing the integration
variable in Eq.~(\ref{cdef}) to $\bs{r}_{13}$, expressing the
integrand as
$\bs{c}_+(\bs{R}_{13}\!+\!\bs{r}_{13}/2,\bs{R}_{13}\!-\!\bs{r}_{13}/2)$,
and expanding $\bs{R}_{13}=\bs{r}_1-\bs{r}_{13}/2$ in $\bs{r}_{13}$ such that
\[
\bs{c}_+(\bs{R}_{13}\!+\!\bs{r}_{13}/2,\bs{R}_{13}\!-\!\bs{r}_{13}/2)
\approx
\bs{c}_+(\bs{r}_1\!+\!\bs{r}_{13}/2,\bs{r}_1\!-\!\bs{r}_{13}/2)
-
\frac{1}{2} \bs{r}_{13}\!\cdot\!\bs{\nabla}_1 
\bs{c}_+(\bs{r}_1\!+\!\bs{r}_{13}/2,\bs{r}_1\!-\!\bs{r}_{13}/2),
\]
where the contribution of the first term vanishes due to the
asymmetry.

\subsection{Summary}

To summarize this section, the continuity equation for the momentum
density (\ref{mconeq}) can be derived from the microscopic Keldysh
approach (within a $\Phi$-derivable approximation) by assuming ``local
uniformity'' within physically small volumes \cite{schwing} (or the
gradient approximation, see below). For short-ranged interactions this
assumption is clearly compatible with the hydrodynamic approach where
one is interested in long-wavelength properties of macroscopic
observables (ideally, orders of magnitude longer than any microscopic
scale \cite{dau6,dau10}). On the other hand, for truly long-ranged
interactions a local stress tensor cannot be constructed leaving only
the integral manifestation of momentum conservation.

\section{Energy conservation}

The Hamiltonian (\ref{H}) does not explicitly depend on time and hence
is invariant under time translations. Hence, energy is conserved and
one should be able to express this fact by means of a continuity
equation
\begin{equation}
\label{coneqen}
\frac{\partial n_E}{\partial t} + \bs{\nabla}\!\cdot\!\bs{j}_E = - \bs{j}\!\cdot\!\bs{\nabla} U,
\end{equation}
where $n_E$ is the energy density and $\bs{j}_E$ is the energy
current. Strictly speaking, Eq.~(\ref{coneqen}) is only exact for the
case of local interactions, similarly to the case of momentum
conservation.

Collisions between neutral molecules described by the traditional
kinetic theory are typically assumed to be local and hence it is not
surprising that Eq.~(\ref{coneqen}) can be straightforwardly obtained
by integrating the kinetic equation multiplied by energy
\cite{dau10}. At the same time, the kinetic theory description of
plasma (i.e., a system -- or gas -- of charged particles) is only
approximate and can be justified at high enough temperatures exceeding
the average interaction energy or at high enough densities where the
average interparticle distance is much smaller than the typical
screening radius \cite{dau10}. The separation between the two length
scales in the problem allows one to distinguish between ``collisions''
-- i.e. the short-distance scattering processes leading to
equilibration and hence contributing to the collision integral -- and
collective phenomena (involving distances of the order of the
screening length) that form the macroscopic fields responsible for the
effective Lorentz force appearing in the left-hand side (LHS) of the
kinetic equation.

\subsection{Global energy conservation}

At the microscopic level, one can account for energy conservation by
considering time translations. Differentiating the Dyson equations
(\ref{dyeq12}) and (\ref{dyeq12cc}) with respect to time and
evaluating their {\it difference}, one finds
\begin{eqnarray}
\label{ec1}
&&
\left[i\frac{\partial}{\partial t_2} \!
\left(i\frac{\partial}{\partial t_1} \!-\! \widehat{\cal H}^{(0)}_1\right)\!
-i\frac{\partial}{\partial t_1} \!
\left(i\frac{\partial}{\partial t_2} \!+\! \widehat{\cal H}^{(0),\dagger}_2\right)\!\right]\! G^{12}_{1,2}=
\\
&&
\nonumber\\
&&
\qquad\qquad\qquad\qquad
=
i\!\int\! d3 \frac{\partial}{\partial t_2} \!\left[ 
\Sigma^R_{1,3} G^{12}_{3,2} \!+\! \Sigma^{12}_{1,3} G^A_{3,2}
\right]
+ 
i\!\int\! d3 \frac{\partial}{\partial t_1} \!\left[ 
G^R_{1,3} \Sigma^{12}_{3,2} \!+\! G^{12}_{1,3} \Sigma^A_{3,2} 
\right]\!.
\nonumber
\end{eqnarray}
Consider now the limit $2\rightarrow1$. In the LHS of Eq.~(\ref{ec1})
one immediately notices
\[
\left.
\left[-\frac{\partial}{\partial t_2}\frac{\partial}{\partial t_1}
+\frac{\partial}{\partial t_1}\frac{\partial}{\partial t_2}\right]
G^{12}_{1,2} \right|_{2\rightarrow1}=0,
\]
and
\begin{equation}
\label{jh}
\left.
\left[-i\frac{\partial}{\partial t_2} U_1 -i\frac{\partial}{\partial t_1}U_2
\right]
G^{12}_{1,2} \right|_{2\rightarrow1}=
U_1\frac{\partial n_1}{\partial t_1}
=
- U_1 \bs{\nabla}_1\!\cdot\!\bs{j}_1,
\end{equation}
where the continuity equation (\ref{coneq}) was used in the last
step. The remaining term in the LHS contains exactly the same
operators as for a non-interacting system and hence can be brought to a form
\begin{equation}
\label{ke0}
\left.\left[-i\frac{\partial}{\partial t_2}\widehat{\cal K}_1
-i\frac{\partial}{\partial t_1}\widehat{\cal K}^\dagger_2\right]\! G^{12}_{1,2}
\right|_{2\rightarrow1} 
= 
\bs{\nabla}_1\!\cdot\!\bs{\cal J}_1 + \frac{\partial {\cal E}^{(1)}_1}{\partial t_1}.
\end{equation}
Here ${\cal E}^{(1)}$ is the ``one-particle'' (``kinetic'') energy
density and $\bs{\cal J}$ is the corresponding current. Specific
expressions for these quantities are determined by the quasiparticle
spectrum and are easier to establish in each particular case. Note,
that the current $\bs{\cal J}$ does contain an explicit interaction
contribution, see below.  Using Eq.~(\ref{ke0}), I may re-write
Eq.~(\ref{ec1}) in the limit $2\rightarrow1$ as
\begin{subequations}
\begin{equation}
\label{ec2}
\frac{\partial {\cal E}^{(1)}_1}{\partial t_1} + \bs{\nabla}_1\!\cdot\!\bs{\cal J}_1
- U_1 \bs{\nabla}_1\!\cdot\!\bs{j}_1 = \Upsilon_1,
\end{equation}
where the RHS is denoted by 
\begin{equation}
\label{ups}
\Upsilon_1 = i\!\left.\left[ 
\frac{\partial}{\partial t_2} \!\int\! d3 \, \Xi(1,2; 3)
\!-\! \frac{\partial}{\partial t_1} \!\int\! d3 \, \Xi^*(2,1; 3) \right] \right|_{2\rightarrow1} .
\end{equation}
\end{subequations}
This quantity is related to the interaction potential, see
Eq.~(\ref{hintid}). Therefore, it is tempting to associate it with the
interaction energy density. This way the volume integral of
Eq.~(\ref{ec2}) yields the integral manifestation of energy
conservation
\begin{equation}
\label{econ}
\frac{\partial}{\partial t} 
\left( \int\!d^dr \, {\cal E}^{(1)} + E_{int} \right)
=
-
\!\int\! d^dr \, \bs{j}\!\cdot\!\bs{\nabla}U.
\end{equation}

\subsection{$\Phi$-derivable approximation and time translations}

To prove the relation between the RHS of Eq.~(\ref{ec1}) and the
interaction energy,consider a change of
the time variable on the ``upper'' branch of the Keldysh contour
\begin{equation}
\label{tt}
t\rightarrow\theta(t)=t+\varphi(t).
\end{equation}
Together with the change of variable, the Green's function acquires an
additional factor \cite{baym}
\begin{equation}
\label{gt3}
\check G_{2,1} \rightarrow \check U(t_2) \check G_{2,1} \check U(t_1),
\quad
\check U = 
\begin{pmatrix}
(\partial\theta/\partial t)^{1/4} & 0 \cr
0 & 1
\end{pmatrix},
\end{equation}
which is needed to cancel the Jacobian in Eq.~(\ref{s0}) so that the
functional (\ref{phidef}) remains invariant. Indeed, every time
integration in any diagram for the functional (\ref{phidef}) involves
four Green's function and hence to cancel the Jacobian, each of them
has to be corrected by a factor of $(\partial\theta/\partial t)^{1/4}$
\cite{baym}. Expanding now in the small variation, one finds similarly
to Eq.~(\ref{gt1})
\begin{eqnarray}
\label{gt3d}
\delta G^{ij}_{2,1} = \delta^{i1}\left[
\frac{\varphi'(t_2)}{4}\!+\!\varphi(t_2)\frac{\partial}{\partial t_2}\right]
G^{ij}_{2,1} 
+
\delta^{j1}\left[
\frac{\varphi'(t_1)}{4}\!+\!\varphi(t_1)\frac{\partial}{\partial t_1}\right]
G^{ij}_{2,1}.
\end{eqnarray}
The expression (\ref{gt3d}) should now be substituted into the
variation (\ref{dphi}) of the functional $\Phi$. This yields
\begin{eqnarray}
\label{dphie1}
&&
\delta\Phi = -
\frac{1}{4}\!\int\! d1 \,\dot\varphi(t_1) \!\int\! d2 
\Big[
\Sigma^{11}_{1,2} G^{11}_{2,1}
\!-\!
\Sigma^{12}_{1,2} G^{21}_{2,1}
\!+\!
G^{11}_{1,2}\Sigma^{11}_{2,1} 
\!-\!
 G^{12}_{1,2}\Sigma^{21}_{2,1}
\Big]
\\
&&
\nonumber\\
&&
\qquad\qquad\qquad\qquad
+
\!\int\! d1 \, \varphi(t_1) \!\int\! d2 
\left[
\Sigma^{11}_{1,2}\frac{\partial}{\partial t_1} G^{11}_{2,1}
\!-\!
\Sigma^{12}_{1,2}\frac{\partial}{\partial t_1} G^{21}_{2,1}
+
\Sigma^{11}_{2,1}\frac{\partial}{\partial t_1} G^{11}_{1,2}
\!-\!
\Sigma^{21}_{2,1}\frac{\partial}{\partial t_1} G^{12}_{1,2}
\right]\!.
\nonumber
\end{eqnarray}
Since the time translation leaves the functional invariant, the above
expression has to be set to zero, $\delta\Phi=0$. 

The first term in Eq.~(\ref{dphie1}) can be re-written with the help of
Eqs.~(\ref{hintid}), (\ref{sigg}), and (\ref{grgr}) as 
\[
-i\!\int\! dt_1 \,\dot\varphi(t_1)
E^{int},
\]
where
\begin{equation}
\label{eint}
E^{int} = \frac{1}{2}\int\! d^d r_1\!\int\! d^d r' \!
\left\langle\hat\psi_H^\dagger(\bs{r}_1,t_1) \hat\psi_H^\dagger(\bs{r}',t_1) 
V(\bs{r}_1\!-\!\bs{r}') 
\hat\psi_H(\bs{r}',t_1) \hat\psi_H(\bs{r}_1,t_1)\right\rangle\!.
\end{equation}
is the interaction energy of the system. Again, using
Eqs.~(\ref{sigg}) and (\ref{grgr}) one can identify the second term in
Eq.~(\ref{dphie1}) with the quantity $\Upsilon_1$, see
Eq.~(\ref{ups}). This way the variation of $\Phi$ takes the form
\begin{equation}
\label{dphie2}
\delta\Phi = - i
\int dt_1 \,\dot\varphi(t_1) E^{int}(t_1)
- i \int d1 \,\varphi(t_1) \Upsilon_1 = 0.
\end{equation}
Integrating the first term by parts and using arbitrariness of
$\varphi(t_1)$, one arrives at the identity
\begin{eqnarray}
\label{id3}
\frac{\partial}{\partial t} E^{int}
+ 
\int d^dr \Upsilon=0,
\end{eqnarray}
which is a manifestation of energy conservation and the justification
for Eq.~(\ref{econ}).

\subsection{Energy conservation in the local approximation}

I now construct the local expression of energy conservation using the
same ``local uniformity'' approximation as in the case of momentum
conservation. Starting with Eq.~(\ref{ec2}), one should notice that
its LHS contains the time derivative of the one-particle
energy density only. To arrive at the total energy density one has to
separate the ``interaction energy'' density from the RHS. To this end,
let me re-write $\Upsilon$ with the help of Eq.~(\ref{hintid}) as
\begin{eqnarray}
\label{upsvint}
&&
\Upsilon_1 
=
-\left.\frac{\partial}{\partial t_2} 
\!\int\! d^dr_3
\left\langle\hat\psi^\dagger(\bs{r}_2, t_2) \hat\psi^\dagger(\bs{r}_3,t_1) 
V(\bs{r}_1\!-\!\bs{r}_3) 
\hat\psi(\bs{r}_3,t_1) \hat\psi(\bs{r}_1, t_1)\right\rangle\right|_{2\rightarrow1}
\\
&&
\nonumber\\
&&
\qquad\qquad\qquad\qquad
-\left.\frac{\partial}{\partial t_1} 
\!\int\! d^dr_3
\left\langle\hat\psi^\dagger(\bs{r}_2, t_2) \hat\psi^\dagger(\bs{r}_3,t_2) 
V(\bs{r}_1\!-\!\bs{r}_3) 
\hat\psi(\bs{r}_3,t_2) \hat\psi(\bs{r}_1, t_1)\right\rangle\right|_{2\rightarrow1}.
\nonumber
\end{eqnarray}
Singling out the interaction energy, see Eq.~(\ref{eint}), this can be
brought to the following form
\begin{eqnarray}
\label{upsvint2}
&&
\Upsilon_1 
=
-\frac{1}{2}\frac{\partial}{\partial t_1} 
\!\int\! d^dr_3
\left\langle\hat\psi^\dagger(\bs{r}_1, t_1) \hat\psi^\dagger(\bs{r}_3,t_1) 
V(\bs{r}_1\!-\!\bs{r}_3) 
\hat\psi(\bs{r}_3,t_1) \hat\psi(\bs{r}_1, t_1)\right\rangle
\\
&&
\nonumber\\
&&
\qquad\qquad\qquad\qquad
+ \frac{1}{2}\!\int\! d^dr_3V(\bs{r}_1\!-\!\bs{r}_3) 
\left\langle\hat\psi^\dagger(\bs{r}_1, t_1)
\frac{\partial}{\partial t_1} \left[ \hat\psi^\dagger(\bs{r}_3,t_1) \hat\psi(\bs{r}_3,t_1) \right]
\hat\psi(\bs{r}_1, t_1)\right\rangle
\nonumber\\
&&
\nonumber\\
&&
\qquad\qquad\qquad\qquad
- \frac{1}{2}\!\int\! d^dr_3V(\bs{r}_1\!-\!\bs{r}_3) 
\left\langle\hat\psi^\dagger(\bs{r}_3, t_1)
\frac{\partial}{\partial t_1} \left[ \hat\psi^\dagger(\bs{r}_1,t_1) \hat\psi(\bs{r}_1,t_1) \right]
\hat\psi(\bs{r}_3, t_1)\right\rangle.
\nonumber
\end{eqnarray}
Introducing the interaction energy density ${\cal E}^{int}$ in the
first term and using the operator relation (\ref{eqmon}) in the last
two, I find
\begin{eqnarray}
\label{upsvint3}
&&
\Upsilon_1 
=
-\frac{\partial{\cal E}_{int}}{\partial t_1} 
-\frac{1}{2}\!\int\! d^dr_3V(\bs{r}_1\!-\!\bs{r}_3) 
\left\langle\hat\psi^\dagger(\bs{r}_1, t_1) \bs{\nabla}_3\hat{\bs{j}}(\bs{r}_3,t_1)
\hat\psi(\bs{r}_1, t_1)\right\rangle
\nonumber\\
&&
\nonumber\\
&&
\qquad\qquad\qquad\qquad
+
\frac{1}{2}\!\int\! d^dr_3V(\bs{r}_1\!-\!\bs{r}_3) 
\left\langle\hat\psi^\dagger(\bs{r}_3, t_1) \bs{\nabla}_1\hat{\bs{j}}(\bs{r}_1,t_1)
\hat\psi(\bs{r}_3, t_1)\right\rangle.
\end{eqnarray}
Substituting the above expression into Eq.~(\ref{ec2}) one can now combine
the time derivative terms into the derivative of the total energy
density, ${n_E={\cal E}^{(1)}+{\cal E}^{int}}$.

To proceed further one need to specify the gradient term in
Eq.~(\ref{ec2}). Without loss of generality, I can write the
``kinetic'' part of the Hamiltonian (\ref{H}) as
\begin{equation}
\label{K}
\widehat{\cal K} = \bs{\nabla}\!\cdot\!\widehat{\bs{K}},
\end{equation}
where the vector $\widehat{\bs{K}}$ carries the dependence on all
additional quantum numbers. The rational for Eq.~(\ref{K}) is the
following. Kinetic energy describes motion and hence the corresponding
operator must not commute with the coordinate. Therefore,
$\widehat{\cal K}$ must be a functional of the gradient operator,
$\bs{\nabla}$, and moreover the formal Taylor series in $\bs{\nabla}$
must start with the first power. Thus each term in the series is
proportional to $\bs{\nabla}$ leading to Eq.~(\ref{K}). Note that
$\widehat{\bs{K}}$ may further depend on $\bs{\nabla}$ as in the
case of the usual parabolic spectrum where
${\widehat{\bs{K}}=-\bs{\nabla}/(2m)}$, while for Dirac fermions in
graphene ${\widehat{\bs{K}}=-iv_g\bs{\sigma}}$, where $v_g$ is the
Fermi velocity and $\bs{\sigma}$ is the vector of the Pauli matrices
\cite{kats}.

Using Eq.~(\ref{K}), one can arrive at the ``explicit'' expressions
for ${\cal E}^{(1)}$ and $\bs{\cal J}$ by evaluating the LHS of
Eq.~(\ref{ke0}). These read
\begin{equation}
\label{ke1}
{\cal E}^{(1)}_1 = -i \left.\bs{\nabla}_1\!\cdot\!\widehat{\bs{K}}_2 G^{12}_{1,2} \right|_{2\rightarrow1},
\qquad
\bs{\cal J}_1 = \left. i
\left(\frac{\partial}{\partial t_1} \widehat{\bs{K}}_2 + 
\frac{\partial}{\partial t_2} \widehat{\bs{K}}_1 \right) G^{12}_{1,2} \right|_{2\rightarrow1}.
\end{equation}
Taking into account the explicit form of $G^{12}_{1,2}$ in terms of the
field operators, Eq.~(\ref{g12}) and using the equations of motion
(\ref{eqmo1}) and (\ref{eqmo2}) to remove the time derivative from
$\bs{\cal J}_\epsilon$, one finds following expression
\begin{equation}
\label{je01}
\bs{\cal J}_1 = \bs{j}^{(1)}_{\epsilon}(1) +  U_1 \bs{j} 
+ \!\int\! d^dr_3V(\bs{r}_1\!-\!\bs{r}_3) 
\left\langle\hat\psi^\dagger(\bs{r}_3, t_1) \bs{\nabla}_1\hat{\bs{j}}(\bs{r}_1,t_1)
\hat\psi(\bs{r}_3, t_1)\right\rangle,
\end{equation}
where the one-particle contribution to the energy current is given by
\begin{equation}
\label{je1}
\bs{j}^{(1)}_{\epsilon}(1) = \left.\left[(\bs{\nabla}_2\!\cdot\!\widehat{\bs{K}}_2) \widehat{\bs{K}}_1
- (\bs{\nabla}_1\!\cdot\!\widehat{\bs{K}}_1) \widehat{\bs{K}}_2\right] G^{12}_{1,2} \right|_{2\rightarrow1}.
\end{equation}

Substituting all of the above results in Eq.~(\ref{ec2}) yields
\begin{equation}
\label{ec3}
\frac{\partial n_E}{\partial t} + \bs{\nabla}\!\cdot\!\bs{j}^{(1)}_{\epsilon}
= - \bs{j}\!\cdot\!\bs{\nabla}U  + \tilde\Upsilon,
\end{equation}
where
\begin{eqnarray*}
&&
\tilde\Upsilon_1 
=
-\frac{1}{2}\!\int\! d^dr_3V(\bs{r}_1\!-\!\bs{r}_3) 
\left\langle\hat\psi^\dagger(\bs{r}_1, t_1) \bs{\nabla}_3\hat{\bs{j}}(\bs{r}_3,t_1)
\hat\psi(\bs{r}_1, t_1)\right\rangle
\nonumber\\
&&
\nonumber\\
&&
\qquad\qquad\qquad\qquad
+
\frac{1}{2}\!\int\! d^dr_3V(\bs{r}_1\!-\!\bs{r}_3) 
\left\langle\hat\psi^\dagger(\bs{r}_3, t_1) \bs{\nabla}_1\hat{\bs{j}}(\bs{r}_1,t_1)
\hat\psi(\bs{r}_3, t_1)\right\rangle
\nonumber\\
&&
\nonumber\\
&&
\qquad\qquad\qquad\qquad
-
\bs{\nabla}_1 \!\int\! d^dr_3V(\bs{r}_1\!-\!\bs{r}_3) 
\left\langle\hat\psi^\dagger(\bs{r}_3, t_1) \hat{\bs{j}}(\bs{r}_1,t_1)
\hat\psi(\bs{r}_3, t_1)\right\rangle,
\end{eqnarray*}
which can be re-written in a more symmetric form
\begin{eqnarray}
\label{ups2}
&&
\tilde\Upsilon_1 
=
-
\frac{1}{2}\bs{\nabla}_1 \!\int\! d^dr_3V(\bs{r}_1\!-\!\bs{r}_3) 
\left\langle\hat\psi^\dagger(\bs{r}_3, t_1) \hat{\bs{j}}(\bs{r}_1,t_1)
\hat\psi(\bs{r}_3, t_1)\right\rangle
\nonumber\\
&&
\nonumber\\
&&
\qquad\qquad\qquad\qquad
-\frac{1}{2}\!\int\! d^dr_3\left[\bs{\nabla}_1V(\bs{r}_1\!-\!\bs{r}_3) \right]
\left\langle\hat\psi^\dagger(\bs{r}_1, t_1) \hat{\bs{j}}(\bs{r}_3,t_1)
\hat\psi(\bs{r}_1, t_1)\right\rangle
\nonumber\\
&&
\nonumber\\
&&
\qquad\qquad\qquad\qquad
-
\frac{1}{2}\!\int\! d^dr_3\left[\bs{\nabla}_1V(\bs{r}_1\!-\!\bs{r}_3) \right]
\left\langle\hat\psi^\dagger(\bs{r}_3, t_1) \hat{\bs{j}}(\bs{r}_1,t_1)
\hat\psi(\bs{r}_3, t_1)\right\rangle.
\end{eqnarray}
The last two terms in Eq.~(\ref{ups2}) are not gradients, but can be
brought to a gradient form in the ``local uniformity'' approximation
\cite{schwing} used above in the case of momentum conservation.
Repeating the steps leading to Eq.~(\ref{piint}), I arrive at the
continuity equation for the energy density (\ref{coneqen}), where the
energy current is defined as
\begin{eqnarray}
\label{jen}
&&
\bs{j}_E = \bs{j}^{(1)}_{\epsilon}
+
\frac{1}{2}\!\int\! d^dr_3V(\bs{r}_1\!-\!\bs{r}_3) 
\left\langle\hat\psi^\dagger(\bs{r}_3, t_1) \hat{\bs{j}}(\bs{r}_1,t_1)
\hat\psi(\bs{r}_3, t_1)\right\rangle
\\
&&
\nonumber\\
&&
\qquad\qquad
-
\frac{1}{4} \!\int\! d^dr_{13} \bs{r}_{13} \left[\nabla^i_{13}V(\bs{r}_1\!-\!\bs{r}_3) \right]
\left[
\left\langle\hat\psi^\dagger(\bs{r}_1\!+\!\bs{r}_{13}/2, t_1) \hat{j}^i(\bs{r}_1\!-\!\bs{r}_{13}/2,t_1)
\hat\psi(\bs{r}_1\!+\!\bs{r}_{13}/2, t_1)\right\rangle\right.
\nonumber\\
&&
\nonumber\\
&&
\qquad\qquad\qquad\qquad\qquad\qquad\qquad\qquad\qquad
+
\left.
\left\langle\hat\psi^\dagger(\bs{r}_1\!-\!\bs{r}_{13}/2, t_1) \hat{j}^i(\bs{r}_1\!+\!\bs{r}_{13}/2,t_1)
\hat\psi(\bs{r}_1\!-\!\bs{r}_{13}/2, t_1)\right\rangle
\right].
\nonumber
\end{eqnarray}

\subsection{Summary}

In this section I have derived the continuity equation for the energy
density (\ref{coneqen}) from the microscopic Keldysh approach by
assuming ``local uniformity'' within small physical volumes
\cite{schwing}. In that sense, the above considerations closely follow
the arguments presented in the previous section. The difference here
is that the resulting expression for the energy current is expressed
in terms of a two-particle Green's function. While it is possible to
express the energy current in terms of the derivatives of the
self-energy and the single-particle Green's function using
Eq.~(\ref{hintid}), the resulting expression is somewhat
cumbersome. At the same time, the obtained expression (\ref{jen}) is
not immediately related to the kinetic equation since the RHS of the
integrated Kadanoff-Baym equation, the quantity $\Upsilon$, has to be
split into the time derivative of the interaction energy and a
contribution towards the energy current. This point will be discussed
in the subsequent section.

\section{Kinetic equation}
\label{sec:kinur}

The local continuity equations Eqs.~(\ref{coneq}), (\ref{mconeq}), and
(\ref{coneqen}) expressing conservation of the number of particles,
momentum, and energy, respectively can be straightforwardly obtained
by integrating the Boltzmann kinetic equation \cite{dau10}. As direct
consequence of the conservation laws, the collision integral vanishes
upon integration. In this section, I discuss the relation of this
property to the identities Eqs.~(\ref{id1}), (\ref{id2}), and
(\ref{id3}).

\subsection{Kadanoff-Baym equation}

Derivation of the kinetic equation is outlined in Ref.~\cite{rammer}.
The calculation is very similar to at least some of the above
considerations, with a few notable points that should be clarified
here. Firstly, Ref.~\cite{rammer} begins with a Dyson's equation for
the Keldysh function $G^K$ instead of Eq.~(\ref{dyeq120}). Combining
it with the conjugate equation one finds the analogue of
Eq.~(\ref{dyeq12dif}), which has the form
\begin{eqnarray}
\label{dyeqKdif}
\left[\left(i\frac{\partial}{\partial t_1} \!-\! \widehat{\cal H}_0(1)\right)
+\left(i\frac{\partial}{\partial t_2} \!+\! \widehat{\cal H}_0(2)\right)\right]\! G^{K}_{1,2}=
\!\int\! d3 \left[ 
\Sigma^R_{1,3} G^{K}_{3,2} + \Sigma^{K}_{1,3} G^A_{3,2}-
G^R_{1,3} \Sigma^{K}_{3,2} - G^{K}_{1,3} \Sigma^A_{3,2} 
\right]\!.
\end{eqnarray}
Now, the Keldysh function is related to the function $G^{12}$ by the identity
\begin{equation}
\label{g12k}
G^K = 2 G^{12} - i A,
\end{equation}
where $A$ is the spectral function defined in Eq.~(\ref{adef}). The
choice of the Keldysh functon as the ``basis'' function for the
nonequilibrium transport theory [as opposed to $G^{12}$ which, after all,
  defines the particle density, see Eq.~(\ref{ndef})] is justified by
the fact \cite{rammer}, that the spectral function does not depend on
the state of the system and hence its contribution to macroscopic
quantities out of equilibrium is irrelevant.

Although Eq.~(\ref{dyeqKdif}) is distinct from Eq.~(\ref{dyeq12dif}),
it has the same property [see Eq.~(\ref{id10})]: the RHS of
Eq.~(\ref{dyeqKdif}) vanishes in the limit $2\rightarrow1$. Indeed,
combining Eqs.~(\ref{dyeq12dif}) and (\ref{dyeqKdif}) according to
Eq.~(\ref{g12k}), one finds the equation for the spectral function
\begin{subequations}
\label{dyeqAdif}
\begin{eqnarray}
\left[\left(i\frac{\partial}{\partial t_1} \!-\! \widehat{\cal H}^{(0)}_1\right)
+\left(i\frac{\partial}{\partial t_2} \!+\! \widehat{\cal H}^{(0)}_2\right)\right]\! A_{1,2}=
\!\int\! d3 \left[ 
\Sigma^R_{1,3} A_{3,2} + \Gamma_{1,3} G^A_{3,2}-
G^R_{1,3} \Gamma_{3,2} - A_{1,3} \Sigma^A_{3,2} 
\right]\!,
\end{eqnarray}
where $\Gamma$ defined in Eq.~(\ref{gdef}) is the self-energy
component analogous to the spectral function. Substituting the
definitions (\ref{adef}) and (\ref{gdef}) into Eq.~(\ref{dyeqAdif}),
one finds for the RHS
\begin{equation}
\label{rhsdiff}
i\!\int\! d3 \left[ 
\Sigma^R_{1,3} G^R_{3,2} - \Sigma^A_{1,3} G^A_{3,2}-
G^R_{1,3} \Sigma^R_{3,2} + G^A_{1,3} \Sigma^A_{3,2} 
\right]\!.
\end{equation}
\end{subequations}
This expression vanishes in the limit $2\rightarrow1$, see
Eq.~(\ref{grgr}), which proves that the RHS of Eq.~(\ref{dyeqKdif})
vanishes in that limit as well.

Finally, one can re-write Eq.~(\ref{dyeqKdif}) introducing the
quantities $A$ and $\Gamma$ in the RHS \cite{rammer}. Introducing the
short-hand notations
\begin{subequations}
\label{notations1}
\begin{equation}
\widehat{\cal D}_{12} = i\frac{\partial}{\partial t_1} +i\frac{\partial}{\partial t_2} 
\!-\! \widehat{\cal H}^{(0)}_1
\!+\! \widehat{\cal H}^{(0)}_2,
\end{equation}
\begin{equation}
(A\otimes B)_{1,2} = \int d3 \, A_{1,3} B_{3,2},
\end{equation}
\begin{equation}
\left[A \,\overset{\otimes}{,}\, B \right]_- = A\otimes B - B\otimes A,
\qquad
\left\{A \,\overset{\otimes}{,}\, B \right\}_+ = A\otimes B + B\otimes A,
\end{equation}
\end{subequations}
one arrives at the variant of the Kadanoff-Baym equation
\begin{eqnarray}
\label{kbeq}
\widehat{\cal D}_{12} G^{K}_{1,2} = \left[{\rm Re}\Sigma \,\overset{\otimes}{,}\, G^K \right]_-
+ \left[ \Sigma^K \,\overset{\otimes}{,}\, {\rm Re} G^R \right]_-
+ \frac{i}{2} \left\{ \Sigma^K \,\overset{\otimes}{,}\, A \right\}_+
- \frac{i}{2} \left\{ \Gamma \,\overset{\otimes}{,}\, G^K \right\}_+.
\end{eqnarray}
This equation is equivalent to Eq.~(\ref{dyeqKdif}) and hence the RHS
vanishes in the limit $2\rightarrow1$. The RHS of the Kadanoff-Baym
equation (\ref{dyeq12dif}) can be brought to the same form as indicated
above.

\subsection{Kinetic equation}

Let me briefly recall the standard steps of the derivation of the
kinetic equation. This can be done in two different ways
\cite{rammer}.

\subsubsection{Quasiparticle and quasiclassical approximations}

The first idea is to apply the gradient approximation to the
Kadanoff-Baym equation (\ref{kbeq}). To do that one first introduces
the Wigner representation (using the relative and center of mass
coordinates introduced in section~\ref{sec:momloc}), see~\ref{appa}.
The Wigner representation is very physical, but unfortunately yields a
complicated expression for the convolution, the so-called Moyal
product \cite{moyal}
\begin{equation}
\label{conv}
A\otimes B = e^{i 
\left(\partial_\epsilon^A\partial_t^B-\bs{\nabla}_{\bs{p}}^A\!\cdot\!\bs{\nabla}_{\bs{r}}^B
-\partial_t^A\partial_\epsilon^B+\bs{\nabla}_{\bs{r}}^A\!\cdot\!\bs{\nabla}_{\bs{p}}^B\right)}
A B.
\end{equation}
The gradient approximation in the Kadanoff-Baym equation is achieved
by keeping the first two terms in the Taylor series for the
exponential in Eq.~(\ref{conv}), which yields
\begin{equation}
\label{grapp}
\left[A \,\overset{\otimes}{,}\, B \right]_- = i \left[A, B\right]_p,
\qquad
\left\{A \,\overset{\otimes}{,}\, B \right\}_+ = 2A B,
\end{equation}
where
\begin{equation}
\label{pb}
\left[A, B\right]_p = \left(\partial_\epsilon A\right)\left(\partial_t B\right) 
-\left(\bs{\nabla}_{\bs{p}}B\right)\!\cdot\!\left(\bs{\nabla}_{\bs{r}}B\right)
-\left(\partial_tA\right)\left(\partial_\epsilon B\right)
+\left(\bs{\nabla}_{\bs{r}}A\right)\!\cdot\!\left(\bs{\nabla}_{\bs{p}}B\right).
\end{equation}
Applying the above approximation to Eq.~(\ref{kbeq}) one finds
\begin{equation}
\label{kbeqw}
\left[ \left(\epsilon-\xi_{\bs{p}}-U-{\rm Re}\Sigma\right), G^K \right]_p -
\left[ \Sigma^K, {\rm Re} G_R\right]_p =  \Sigma^K A - \Gamma G^K.
\end{equation}
Here all the derivatives are combined in the LHS, while the
remaining RHS can be identified with the collision integral.

The last step in the derivation is based on a further
approximation. The ``quasiparticle approximation'' relies on the
Kadanoff-Baym solution \cite{bk} for the spectral function which can
be approximated by a $\delta$-function,
$A=2\pi\delta(\epsilon-\xi_{\bs{p}}-U)$. Combined with the
corresponding form of the Keldysh Green's function, $G^K=-2\pi i
h_{\bs{p}} \delta(\epsilon-\xi_{\bs{p}}-U)$, where $h_{\bs{p}}$
defines the conventional distribution function,
$f_{\bs{p}}=(1-h_{\bs{p}})/2$, one integrates Eq.~(\ref{kbeqw}) over
$\epsilon$ and obtains the standard (Boltzmann) kinetic equation. The collision
integral (up to a numerical factor) takes the form $I\propto
\Sigma^K[h_{\bs{p}}] - (\Sigma^R-\Sigma^A) h_{\bs{p}}$ and is the
function of $\bs{p}$, $\bs{r}$, and $t$. Summing over all states now
amounts to integrating over the momentum variable $\bs{p}$. Given that
it appears as a Fourier transform inthe relative coordinate, such
integration is equivalent to the limit $2\rightarrow1$ considered
above.

Alternatively, one can integrate over $\xi_{\bs{p}}$ (the
``quasiclassical approximation''). The idea is that in the case of,
e.g., electro-phonon interaction the self-energy acquires energy
dependence and hence the Kadanoff-Baym solution for the spectral
function can no lnger be reduced to the above
$\delta$-function. However, should the momentum dependence of the
self-energy remain weak (e.g., due to the Migdal theorem
\cite{migdal}) the spectral function retains the form of a sharp peak
in the variable $\xi_{\bs{p}}$. This relies on the existence of the
Fermi surface: upon integration over $\xi_{\bs{p}}$ the Green's
functions are essentially restricted to the Fermi surface and depend
only on the orientation of momentum. While the quasiclassical
approximation does not explicitly require the mixed representation in
the time variables, it is often invoked in order to reach the standard
form of the kinetic equation \cite{rammer}. The procedure becomes
rather similar to the previous case and yields the same form of the
collision integral, where both the self-energies and the distribution
function now depend on $\epsilon$ instead of the absolute value of
$\bs{p}$. Assuming the existence of quasiparticles, the two
approximations can be related by the formal introduction of the
density of states. At the same time, the quasiclassical approximation
does not rely on the quasiparticle paradigm, which formally is
expressed through the fact that the energy and momentum variables are
no longer related. A known limitation of the quasiclassical
approximation is its reliance on the particle-hole symmetry
\cite{rammer} which precludes one from describing, e.g.,
thermoelectric effects.

\subsection{Beyond the quasiclassical approximation}

The second approach to deriving the kinetic equation does not rely on
the quasiparticle approximation. Instead, one introduces the {\it
  Ansatz} \cite{rammer}
\begin{equation}
\label{gka}
G^K=G^R\otimes h - h \otimes G^A.
\end{equation}
Using this form in the Dyson's equation (\ref{dyeqKdif}) and taking
into account the diagonal elements of Eq.~(\ref{dyeqk}), one arrives
at the equation
\begin{equation}
\label{heq}
G^R\otimes {\cal B} - {\cal B} \otimes G^A = 0,
\qquad
{\cal B} = 
\widehat{\cal D}_{12} h - \left[{\rm Re}\Sigma \,\overset{\otimes}{,}\, h \right]_-
+ \frac{i}{2} \left\{ \Gamma \,\overset{\otimes}{,}\, h \right\}_+
+ \Sigma^K .
\end{equation}
Solving this equation to the leading order of the gradient expansion
amounts to setting ${\cal B}=0$ which seemingly yields the same form
of the collision integral, $I\propto \Sigma^K[h] - (\Sigma^R-\Sigma^A)
h$, albeit obtained without any recourse to the quasiparticle
approximation. The difference is that here the ``distribution
function'' depends not only on $\bs{p}$, $\bs{r}$, and $t$ as in the
case of the quasiclassical approximation, but also on the energy
variable, i.e. $h=h(\bs{p},\epsilon; \bs{r}, t)$. The quasiclassical
(or quasiparticle) approximation allows one to integrate over
$\epsilon$ using the ``$\delta$-peak''-like form of the spectral
function. The Ansatz (\ref{gka}) leads to the kinetic equation without
any additional assumptions on the form of $A$.

An alternative method of deriving the quantum kinetic equation was
suggested in Ref.~\cite{ivanov} on the basis of the observation that
all elements of the Keldysh Green's function matrix could be expressed
in terms of two functions only, cf. Eqs.~(\ref{gfcon}) and
(\ref{red1}). Choosing the spectral function as one of the two and
noticing that it becomes real in the Wigner representation, one can
introduce another real function in the Wigner representation,
$h=h(\bs{p},\epsilon; \bs{r}, t)$, such that 
\begin{subequations}
\label{g12a}
\begin{equation}
G^{12}(\bs{p},\epsilon; \bs{r}, t)=iA(\bs{p},\epsilon; \bs{r}, t)h(\bs{p},\epsilon; \bs{r}, t),
\qquad
G^{21}(\bs{p},\epsilon; \bs{r}, t)=-iA(\bs{p},\epsilon; \bs{r}, t)
\left[1-h(\bs{p},\epsilon; \bs{r}, t)\right].
\end{equation}
This allows to use the functions $A$ and $h$ to express the Keldysh
function $G^K$
\begin{equation}
\label{gkw}
G^K(\bs{p},\epsilon; \bs{r}, t)=-iA(\bs{p},\epsilon; \bs{r}, t)
\left[1-2h(\bs{p},\epsilon; \bs{r}, t)\right].
\end{equation}
\end{subequations}
Expressing the self-energies in a similar way
\begin{subequations}
\label{s12a}
\begin{equation}
\Sigma^{12}(\bs{p},\epsilon; \bs{r}, t)=
i\Gamma(\bs{p},\epsilon; \bs{r}, t)\gamma(\bs{p},\epsilon; \bs{r}, t),
\qquad
\Sigma^{21}(\bs{p},\epsilon; \bs{r}, t)=-i\Gamma(\bs{p},\epsilon; \bs{r}, t)
\left[1-\gamma(\bs{p},\epsilon; \bs{r}, t)\right],
\end{equation}
with
\begin{equation}
\label{skw}
\Sigma^K(\bs{p},\epsilon; \bs{r}, t)=-i\Gamma(\bs{p},\epsilon; \bs{r}, t)
\left[1-2\gamma(\bs{p},\epsilon; \bs{r}, t)\right],
\end{equation}
\end{subequations}
one can use the new notations to re-write Eq.~(\ref{kbeq}) as
\begin{eqnarray}
\label{kbeqv}
\widehat{\cal D} Ah - \left[{\rm Re}\Sigma \,\overset{\otimes}{,}\, Ah \right]_p
- \left[ \Gamma\gamma \,\overset{\otimes}{,}\, {\rm Re} G^R \right]_p
= \Gamma A (\gamma - h),
\end{eqnarray}
where the RHS is essentially the same form of the collision integral
expressed in the new notations.

The ``quantum kinetic equation'' (\ref{kbeqv}) was obtained within the
leading order of the gradient approximation and hence provides a
quantitative condition for its validity
\begin{equation}
\label{condgra}
|\gamma-h|\ll1.
\end{equation}
Consequently, in the LHS of Eq.~(\ref{kbeqv}) one can replace $\gamma$
by $h$. The resulting equation corresponds to the Botermans and
Malfliet \cite{boter} choice of the quantum kinetic equation (as opposed
to the original Kadanoff-Baym choice). Both variants are equivalent
within the applicability range of the gradient approximation.

In comparison to the Ansatz (\ref{gka}), the variable choice
(\ref{g12a}) reintroduces the spectral function in the definition of
the distribution function $h$, while leaving the energy dependence of
the latter. On the other hand, the choice (\ref{g12a}) is always
possible \cite{ivanov}, while the Ansatz (\ref{gka}) is guaranteed to
be valid only within the gradient approximation \cite{rammer}.

\subsection{Kadanoff-Baym equation and the continuity equation}

Let me now compare the derivation of the continuity equation presented
in section~\ref{ce1} to the well-known approach of integrating the
kinetic equation. Particle number conservation is manifested in the
traditional kinetic theory by the fact that collision integral
vanishes after being summed up over all states \cite{dau10}. The
quantum kinetic equation, regardless of the variant,
cf. Eqs.~(\ref{kbeqw}), (\ref{gkw}). and (\ref{kbeqv}), contains also
the renormalization terms in the LHS. Consequently, the derivation of
the kinetic equation consists of making sure that the integral of the
collision term vanishes and at the same time that the renormalization
terms do not affect the particle density and current \cite{rammer}.

In contrast, the argument presented in section~\ref{ce1} relies on the
single identity, Eq.~(\ref{id10}), where taking the limit
$2\rightarrow1$ is equivalent to integrating the collision integral
over all energies and momenta, $\epsilon$ and $\bs{p}$. The
combination of the self-energies and Green's functions in
Eq.~(\ref{id10}) comprises both the collision integral and
renormalization terms (before the gradient approximation). However,
vanishing of these terms together does not in general guarantee that
they should vanish individually although this does happen for most
common forms of the kinetic equation \cite{rammer}. The fact that
renormalization does not affect the particle density follows from the
operator definition, Eq.~(\ref{ndef}). Similarly, the current $\bs{j}$
is determined by the operator form of the continuity equation,
Eq.~(\ref{eqmon}) and hence cannot be affected by interaction
explicitly. This does not mean that the density and current in an
interacting system are the same as in non-interacting one: both
definitions involve the {\it exact} Green's function $G^{12}$, which
can be very different from the free-particle one. In that sense,
vanishing of the renormalization terms ensures consistency of
definitions of macroscopic currents and densities in the microscopic
and kinetic theories. Of course, the total number of particles is the
same as in the free system since interaction does not ``produce'' or
``destroy''any particles.

Finally, let me reiterate that the continuity equation (\ref{coneq})
is exact as long as the interaction (and any potential) is expressed
in terms of particle density, as is the case with most typical models
(electron-electron Coulomb interaction, electron-phonon -- or any
other boson -- coupling, electron-impurity scattering, etc). The
purpose of the identity (\ref{id10}) is to make sure that any
approximation made for Green's functions and self-energies does not
violate the conservation law.

\subsection{Kadanoff-Baym equation and momentum conservation}

The continuity equation for momentum density (\ref{mconeq}) is the
central equation in the hydrodynamic theory eventually yielding the
Euler and Navier-Stokes equations. In contrast to the continuity
equation (\ref{coneq}), the equation (\ref{mconeq}) is not exact, but
is valid within the gradient approximation. This is not a problem,
since hydrodynamics describes long-wavelength variations of macroscopic
quantities. The same gradient approximation is used to derive the
kinetic equation. The equation (\ref{mconeq}) can then be obtained by
multiplying the kinetic equation by momentum and summing over all
states without further approximations. As a result of this procedure
the RHS (i.e., the collision integral) of the kinetic equation
vanishes which is the manifestation of momentum conservation
\cite{dau10}.

Microscopically, momentum conservation is manifested through the
identity (\ref{id2}). This directly leads to vanishing of the quantity
$\bs{C}$ integrated over all space and hence to the global (integral)
relation (\ref{momcontot}). The quantity $\bs{C}$ itself emerges from
the RHS of the Kadanoff-Baym equation in the limit $2\rightarrow1$
which is equivalent of integrating the RHS of the quantum kinetic
equation. The gradient approximation used to derive Eq.~(\ref{mconeq})
is equivalent to the one needed to derive the quantum kinetic
equation. In particular, separating the integrand $\bs{c}_+$ into two
parts corresponds to the distinction between the collision integral
and the renormalization terms in Eqs.~(\ref{kbeqw}) and
(\ref{kbeqv}). In this case it can be seen directly that the
integrated collision integral vanishes while the renormalization terms
contribute to the momentum flux tensor, see Eq.~(\ref{piint}). The
momentum density $\bs{g}$ is determined by the momentum operator, see
Eq.~(\ref{mndef}), and hence is unaffected by renormalizations
similarly to the particle number density and current. The derivation
presented in Sec.~\ref{sec:momcon} is thus equivalent to the more
standard route of going through the kinetic equation (either the
Boltzmann one or quantum), but has the advantage of being free of any
additional approximation beyond the gradient expansion. Taking into
account additional interaction that do not conserve momentum amounts
to evaluating its contribution to the quantity $\bs{C}$ in the
``0-th'' approximation with respect to the gradients, which is
equivalent to evaluating the corresponding collision integral.

\subsection{Kadanoff-Baym equation and energy conservation}

Energy conservation is the most difficult part of the presented
approach since the energy density at the operator level is essentially
a two-particle correlation function. Global energy conservation can be
expressed in terms of the integral relation (\ref{econ}). As in the
case of momentum conservation, the RHS of the Kadanoff-Baym equation
in the limit $2\rightarrow1$, i.e., the quantity $\Upsilon$,
determines the time derivative of the interaction energy upon being
integrated over all space. However, now both the energy density and
current are renormalized by interaction. Separating the time
derivative of the interaction energy density from $\Upsilon$ leaves
the contribution to the energy current that has to be combined with
the interaction contribution to the ``single-particle'' current
$\bs{\cal J}$. This should be contrasted with the standard kinetic
theory derivation \cite{dau10,me0,me1} where a direct integration of
the kinetic equation multiplied by energy yields the energy density
and current from the LHS, while the collision integral vanishes. At
the same time, the ``internal energy'' appears though thermodynamic
identities \cite{dau10}. This apparent complication in comparing the
two approaches is reminiscent of the common practice in conventional
hydrodynamics where dissipative processes are taken into account using
the entropy flow equation rather than the continuity equation for the
energy density \cite{dau6,dau10}. The entropy flow equation will be
discussed in a forthcoming publication \cite{menext}.

Recent literature on electronic hydrodynamics in graphene
\cite{rev,luc,me1,me0} devotes little attention to the internal
energy. The role of electron-electron interaction is seen as being
responsible for equilibration, although in real materials
equilibration is most likely to occur with the help of phonons.
Taking into account electron-phonon interaction leading to energy
relaxation \cite{meig1} would violate the identity (\ref{id2}). In the
simplest case (cf. the arguments of Ref.~\cite{meig1}), one would have
to evaluate the phonon contribution to $\Upsilon$ establishing the
weak decay contribution to the continuity equation (\ref{coneqen}).

\section{Discussion}

In this paper I have presented a detailed derivation of the local
continuity equations providing the basis of the hydrodynamic theory of
electronic transport. While the continuity equation manifesting gauge
invariant is exact, the corresponding equations for the momentum and
energy density are obtained within the gradient approximation. The
presented derivation is more general than the kinetic theory approach
since it relies neither on additional approximations (such as the
common quasiparticle or quasiclassical approximations) nor on the
concept of the distribution function. Although the latter can be
introduced at the quantum level [e.g., by Eqs.~(\ref{gka}) or
  (\ref{g12a})], it is not always obvious how to generalize this
quantity to more complicated cases, e.g., involving spin-orbit
interaction. Keeping the discussion in coordinate space allows for a
direct generalization for systems in confined geometries.

The idea that hydrodynamics is ``more general'' than the kinetic
theory is not new and can be already seen in the original hydrodynamic
description of conventional fluids (none of which could be described
by a kinetic equation). Microscopic expressions for the momentum flux
tensor, interaction energy density, and energy current presented here
open a direct pathway for evaluating these quantities using specific
models of the systems of interest. In particular, there is already a
substantial literature on hydrodynamic approach to ``strange'' or
``bad'' metals \cite{zaa14,legros19,har15,erd18,zaa19}, where the
excitation spectrum might not contain usual quasiparticles, as could
be seen in photoemission \cite{arpesrev} and transport \cite{ayres}
experiments. Some of these materials exhibit resistance that is linear
in temperature over a wide range including both low and high
temperatures \cite{bruin}, the behavior that has been puzzling the
community for decades. The analysis presented here could be seen as a
way of evaluating resistivity directly (similarly to the case of
graphene \cite{me0,rev,luc}) without the need for a Kubo formula and
may prove helpful for describing less established systems such as
``non-Fermi liquids''.

Finally, supplementing the continuity equations by the constitutive
relations one arrives at the hydrodynamic theory. The constitutive
relations are typically formulated either on symmetry grounds
\cite{dau6} or on the basis of the kinetic theory
\cite{dau10,me0,me1}. In the latter case one associates the ideal flow
with local equilibrium. All macroscopic quantities (the densities,
currents, and stress tensor) can be straightforwardly evaluated by
substituting the explicit form of the local equilibrium distribution
function into their respective definitions. Dissipation is taken into
account perturbatively insofar the dissipative corrections are
expressed within the leading order in the gradient expansion. The
particular form of the dissipative crrections is dictated by symmetry
leaving a small number of coefficients (in the usual hydrodynamics the
three coefficients are the shear and bulk viscosities and thermal
conductivity) to be determined either by solving the kinetic equation
or phenomenologically.

In the present approach the macroscopic quantities are expressed in
terms of the exact Green's functions. The expression for the fluid
velocity in terms of the Keldysh Green's functions was given in
Ref.~\cite{aa17}. Since the general form of the dissipative
corrections is independent of the microscopic derivation, one needs to
expand Eqs.~(\ref{jd0}), (\ref{piint}), and (\ref{jen}) in terms of
the velocity gradients and hence determine the dissipative
coefficients (the viscosity, electrical and thermal
conductivities). If the calculation is done perturbatively, then the
results are going to coincide with those done within the kinetic
approach (the diagrammatic perturbative expansion is identical with
the kinetic theory \cite{zna}). However, if there are no
quasiparticles in the system such that the kinetic theory breaks down,
the presented approach still offers a straightforward way to evaluate
the kinetic coefficients. Alternatively, one can treat the
coefficients purely phenomenologically while using the hydrodynamic
equations to determine the spatial distribution of the charge and
energy flows.

\section*{Acknowledgments}

The author wishes to thank I.V. Aleiner, I.V. Gornyi, A.D. Mirlin,
J. Schmalian, and A. Shnirman for fruitful discussions. This work was
supported by the German Research Foundation DFG project NA 1114/5-1
and the European Commission under the EU Horizon 2020 MSCA-RISE-2019
Program (Project 873028 HYDROTRONICS).


\appendix

\section{Non-equilibrium (or Keldysh) Green's function formalism}
\label{appa}

Here I summarize the notations for the Keldysh Green's functions and
their standard relations to keep the paper self-complete. For a
detailed account of the Keldysh technique see Refs.~\cite{rammer,kam}.

\subsection{Keldysh Green's function}

The central quantity of the formalism is the Green's function that can
be defined either in the Heisenberg (subscript ``H'') or ``interaction
(subscript ``I'') representation on the Keldysh contour (subscript
``C'')
\begin{subequations}
\label{gfdef0}
\begin{eqnarray}
G(1_C, 2_C) = -i \left\langle {\cal T}_C \hat\psi_H(1_C)\hat\psi^\dagger_H(2_C)\right\rangle
=-i\left\langle {\cal T}_C \widehat{\cal S}_C \hat\psi_I(1_C)\hat\psi^\dagger_I(2_C)\right\rangle,
\end{eqnarray}
where the latter expression retains only the ``connected''
diagrams. In Eq.~(\ref{gfdef0}), ${\cal T}_C$ is the time-ordering
operator on the Keldysh contour and the ``scattering matrix'' is
\begin{equation}
\label{s0}
\widehat{\cal S}_C = {\cal T}_C \exp\left[-i\int\limits_C dt_C \widehat{H}_{\rm int}(t_C)\right].
\end{equation}
\end{subequations}

The Green's function (\ref{gfdef0}) can be more conveniently described
in the matrix form
\begin{equation}
\label{gfm}
\check G_{1,2} =
\begin{pmatrix}
G^{11}_{1,2} & G^{12}_{1,2} \cr
G^{21}_{1,2} & G^{22}_{1,2}
\end{pmatrix},
\end{equation}
where
\begin{subequations}
\label{gfdef}
\begin{equation}
\label{g12}
G^{12}_{1,2} = i\left\langle \hat\psi^\dagger_H(2) \hat\psi_H(1) \right\rangle,
\end{equation}
\begin{equation}
\label{g21}
G^{21}_{1,2} = -i\left\langle \hat\psi_H(1) \hat\psi^\dagger_H(2) \right\rangle,
\end{equation}
\begin{equation}
\label{g11}
G^{11}_{1,2} = \theta(t_1-t_2) G^{21}_{1,2} + \theta(t_2-t_1) G^{12}_{1,2},
\end{equation}
\begin{equation}
\label{g22}
G^{22}_{1,2} = \theta(t_1-t_2) G^{11}_{1,2} + \theta(t_2-t_1) G^{21}_{1,2}.
\end{equation}
\end{subequations}
The four matrix elements are not independent and satisfy
\begin{subequations}
\label{gfcon}
\begin{equation}
\!\!\!\check G_{1,2} = -\check\tau_1 \check G^\dagger_{2,1} \check\tau_1 ,
\;\;
{\rm Tr\,}\check G_{1,2} = {\rm Tr\,}\check\tau_1\check G_{1,2},
\end{equation}
where $\check\tau_i$ are the Pauli matrices in the ``Keldysh space''.
Explicitly, the later relation takes the form
\begin{equation}
\label{gfcon1}
G^{11}_{1,2}+G^{22}_{1,2} = G^{12}_{1,2}+G^{21}_{1,2}.
\end{equation}
\end{subequations}

\subsection{Self-energy}

The Green's function obeys the formally exact Dyson's equation
\begin{equation}
\label{dyeq0}
\left(i\frac{\partial}{\partial t_1} \!-\! \widehat{\cal H}^{(0)}_1\right)\! \check G_{1,2}
-
\!\int\! d3 \, \check \Sigma_{1,3} \check\tau_3 \check G_{3,2}
=
\check\tau_3 \delta_{1,2},
\end{equation}
where  the self-energy is the matrix
\begin{equation}
\label{sem}
\check \Sigma_{1,2} =
\begin{pmatrix}
\Sigma^{11}_{1,2} & \Sigma^{12}_{1,2} \cr
\Sigma^{21}_{1,2} & \Sigma^{22}_{1,2}
\end{pmatrix}.
\end{equation}
The above definition differs from that in Ref.~\cite{dau10},
where the Pauli matrix in the integral in Eq.~(\ref{dyeq0}) precedes
the self-energy \cite{dau10}. This amounts to the replacement
\[
\check \Sigma_{1,2} \rightarrow \check\tau_3\check \Sigma_{1,2} \check\tau_3,
\]
or simply put, the extra minus sign for the off-diagonal
elements. This can be made clearer by transforming the
integro-differential equation (\ref{dyeq0}) to the integral form using
the ``free'' Green's function
\begin{equation}
\label{g0}
\check G^{(0)}_{1,2} = \left(i\frac{\partial}{\partial t_1} \!-\! \widehat{\cal H}^{(0)}_1\right)^{\!-1}
\!\check\tau_3 \delta_{1,2}.
\end{equation}
Applying the operator $\left(i\partial_{t_1}-\widehat{\cal H}^{(0)}_1\right)^{\!-1}$
to Eq.~(\ref{dyeq0}) from the left, one finds
\begin{equation}
\label{dyeq01}
\check G_{1,2} - 
\!\int\! d3d4 \, \check G^{(0)}_{1,4} \check\tau_3\check \Sigma_{4,3} \check\tau_3 \check G_{3,2}
=\check G^{(0)}_{1,2},
\end{equation}
in contrast to the corresponding equation in Ref.~\cite{dau10} where
there are no Pauli matrices.

The rationale for the above notation is as follows. The self-energy
has the same ``symmetry'' as the Green's function, see
Eq.~(\ref{gfcon})
\begin{subequations}
\label{selfcon}
\begin{equation}
\!\!\!\check \Sigma_{1,2} = -\check\tau_1 \check \Sigma^\dagger_{2,1} \check\tau_1 ,
\;\;
{\rm Tr\,}\check \Sigma_{1,2} = {\rm Tr\,}\check\tau_1\check \Sigma_{1,2}.
\end{equation}
The latter relation reads
\begin{equation}
\label{selfcon1}
\Sigma^{11}_{1,2}+\Sigma^{22}_{1,2}=\Sigma^{12}_{1,2}+\Sigma^{21}_{1,2},
\end{equation}
\end{subequations}
similarly to Eq.~(\ref{gfcon1}) and without the extra minus sign in the
right-hand side as in Ref.~\cite{dau10}.

\subsection{Keldysh rotation}

One may try to use the relations (\ref{gfcon}) to reduce the number of
Green's functions. This can be achieved by a ``rotation'' \cite{rammer,larov}
\begin{subequations}
\label{lor}
\begin{equation}
\check G_{1,2} \rightarrow \frac{1}{2} 
\begin{pmatrix}
1 & -1 \cr
1 & 1
\end{pmatrix}
\check \tau_3 \check G_{1,2}
\begin{pmatrix}
1 & 1 \cr
-1 & 1
\end{pmatrix}.
\end{equation}
In the new basis, both the Green's function and self-energy have the
similar form (unlike the form suggested in Ref.~\cite{dau10})
\begin{equation}
\label{gfk}
\check G =
\begin{pmatrix}
G^{R} & G^{K} \cr
0 & G^{A}
\end{pmatrix},
\qquad
\check \Sigma =
\begin{pmatrix}
\Sigma^{R} & \Sigma^{K} \cr
0 & \Sigma^{A}
\end{pmatrix}.
\end{equation}
In terms of the original Green's functions, the newly defined
functions are given by
\begin{eqnarray}
\label{grak}
&&
G^R=G^{11}-G^{12}=G^{21}-G^{22}=\theta(t_1-t_2)\left[G^{21}_{1,2}-G^{12}_{1,2}\right],
\nonumber\\
&&
\nonumber\\
&&
G^A=G^{11}-G^{21}=G^{12}-G^{22}=-\theta(t_2-t_1)\left[G^{21}_{1,2}-G^{12}_{1,2}\right],
\\
&&
\nonumber\\
&&
G^K=G^{12}+G^{21}=G^{11}+G^{22}.
\nonumber
\end{eqnarray}
\end{subequations}
In the rotated basis, the Dyson's equation takes the form (same as in Ref.~\cite{dau10})
\begin{equation}
\label{dyeqk}
\left(i\frac{\partial}{\partial t_1} \!-\! \widehat{\cal H}^{(0)}_1\right)\! \check G_{1,2}
-
\!\int\! d3 \, \check \Sigma_{1,3} \check G_{3,2}
= \delta_{1,2}.
\end{equation}

The basis rotation does not completely eliminate the redundancy in the
definitions of the Green's function. Indeed, the Green's function in
the rotated basis satisfies [cf. Eq.~(\ref{gfcon})]
\begin{equation}
\label{red1}
\check G_{1,2} = \check\tau_2 \check G^\dagger_{2,1} \check\tau_2,
\end{equation}
with the similar constraint on the self-energy [cf. Eq.~(\ref{selfcon})]
\begin{equation}
\label{selfen1}
\check \Sigma_{1,2} = \check\tau_2 \check \Sigma^\dagger_{2,1} \check\tau_2.
\end{equation}
As a result, only two functions are in either matrix are independent.

Suppose one chooses $G^{12}_{1,2}$ and $G^{21}_{1,2}$ as such
independent functions. Then from Eq.~(\ref{gfcon}) it follows that
they have the following property
\begin{equation}
\label{gfcon12}
\left[G^{12}_{1,2}\right]^* = - G^{12}_{2,1},
\quad
\left[G^{21}_{1,2}\right]^* = - G^{21}_{2,1}.
\end{equation}
Consider then their difference
\begin{subequations}
\label{adef}
\begin{equation}
A_{1,2}=i\left[G^{21}_{1,2}-G^{12}_{1,2}\right]=i\left[G^{R}_{1,2}-G^{A}_{1,2}\right].
\end{equation}
As follows from the symmetry of the Green's functions,
Eq.~(\ref{gfcon12}), the new function satisfies
\begin{equation}
\label{acon}
A^*_{1,2}=A_{2,1}.
\end{equation}
\end{subequations}

\subsection{Wigner representation}

The gradient approximation needed to derive quantum kinetic equations
is most readily demonstrated in the mixed or Wigner representation,
i.e. the Fourier representation in the relative coordinate (the
relative and center of mass coordinates were introduced in
section~\ref{sec:momloc})
\begin{equation}
\label{wig}
A(\bs{p},\epsilon; \bs{R}_{12}, T_{12})
=
\!\int\! d^dr_{12} dt_{12} A_{1,2}
e^{-i(\bs{p}\bs{r}_{12}-\epsilon t_{12})},
\end{equation}
where ${t_{12}=t_1-t_2}$ and ${T_{12}=(t_1+t_2)/2}$.

In the Wigner representation, the spectral function $A$
is real [cf. Eq.~(\ref{acon})] and satisfies the ``sum rule''
\begin{equation}
\label{aconw}
A^*(\bs{p},\epsilon; \bs{r}, t)=A(\bs{p},\epsilon; \bs{r}, t),
\qquad
\int\limits_{-\infty}^\infty\!\frac{d\epsilon}{2\pi}A(\bs{p},\epsilon; \bs{r}, t)=1.
\end{equation}
At the same time, it completely determines the retarded and advanced
functions through the relation
\begin{equation}
\label{graw}
G^R(\bs{p},\epsilon; \bs{r}, t)\!=\!\left[G^A(\bs{p},\epsilon; \bs{r}, t)\right]^*
\!=\!
\!\int\limits_{-\infty}^\infty\!\frac{d\epsilon'}{2\pi}
\frac{A(\bs{p},\epsilon'; \bs{r}, t)}{\epsilon-\epsilon'+i0^+},
\end{equation}
where $(\epsilon+i0^+)^{-1}$ is the Fourier transform of the
$\theta$-function in Eq.~(\ref{grak}). As a result, one may express these function as
\begin{equation}
\label{regr}
G^{R(A)} = {\rm Re} \; G^R \mp \frac{i}{2} A.
\end{equation}

Similar relations can be defined for the self-energy. Defining the
analogue of the spectral function
\begin{equation}
\label{gdef}
\Gamma_{1,2}=i\left[\Sigma^{R}_{1,2}-\Sigma^{A}_{1,2}\right],
\end{equation}
one finds in the Wigner representation
\begin{equation}
\label{resig}
\Sigma^{R(A)} = {\rm Re} \Sigma \mp \frac{i}{2} \Gamma.
\end{equation}

\bibliographystyle{elsarticle-num}

\bibliography{hydro-refs,refs-books}

\end{document}